\documentclass[%
 reprint,
 amsmath,amssymb,
 aps
]{revtex4-2}

\usepackage{graphicx}
\usepackage{dcolumn}
\usepackage{bm}

\usepackage{hyperref}
\usepackage{algorithm}
\usepackage{algpseudocode}
\usepackage{xcolor}
\usepackage{upgreek}

\usepackage{hyperref}

\hypersetup{
    colorlinks=true,
    linkcolor=black,
    citecolor=black,
    filecolor=magenta,      
    urlcolor=blue,
    pdfpagemode=FullScreen,
    }

\newcommand{\refeq}[1]{Eq.~(\ref{eq:#1})}          
          
\newcommand{\reffig}[1]{Figure~\ref{fig:#1}}          
\newcommand{\refsec}[1]{Section~\ref{sec:#1}}
\newcommand{\refapp}[1]{Appendix~\ref{app:#1}}
\newcommand{\reftab}[1]{Table~\ref{tab:#1}}
\newcommand{\refalg}[1]{Algorithm~\ref{alg:#1}} 

\begin{document}

\preprint{APS/123-QED}

\title{Mode-by-mode Relative Binning: Fast Likelihood Estimation for Gravitational Waveforms with Spin-Orbit Precession and Multiple Harmonics}

\author{Nathaniel Leslie}
\email{nathaniel\_leslie@berkeley.edu}
\author{Liang Dai}
\email{liangdai@berkeley.edu}
\affiliation{Department of Physics\\
University of California, Berkeley, CA 94720, USA}
\author{Geraint Pratten}
\email{g.pratten@bham.ac.uk}
\affiliation{School of Physics and Astronomy \& Institute for Gravitational Wave Astronomy, University of Birmingham\\ 
Birmingham, B15 2TT, UK}

\date{\today}

\begin{abstract}
Faster likelihood evaluation enhances the efficiency of gravitational wave signal analysis. We present Mode-by-mode Relative Binning (MRB), a new method designed for obtaining fast and accurate likelihoods for advanced waveform models that include spin-orbit precession effects and multiple radiation harmonics from compact binary coalescence. Leveraging the ``twisting-up'' procedure of constructing precessing waveform modes from non-precessing ones, the new method mitigates degrade of relative binning accuracy due to interference from superimposed modes. Additionally, we supplement algorithms for optimizing the choice of frequency bins specific to any given strain signal under analysis. Using the new method, we are able to evaluate the likelihood with up to an order of magnitude reduction in the number of waveform model calls per frequency compared to the previously used relative binning scheme, and achieve better likelihood accuracy than is sufficient for obtaining source parameter posterior distributions that are indistinguishable from the exact ones.
\end{abstract}

\maketitle

\section{Introduction}
\label{sec:intro}
Astrophysical information about gravitational wave sources is commonly extracted via parameter inference within the Bayesian framework. This process requires thorough sampling of a high-dimensional parameter space (e.g. with a total of 10--15 source parameters), and hence a huge number (typically on the order of $10^7$--$10^8$) of likelihood evaluations. To map either the likelihood function or the posterior distribution, it is important to accurately evaluate the likelihood function on manageable timescales.

Brute-force calculation of the likelihood is often costly. In order to be compatible with Fast Fourier Transform (FFT), it requires evaluating the frequency-domain model waveform on a uniformly sampled, sufficiently fine grid that has $\sim T\,f_{\rm max}$ frequencies, where $f_{\rm max}$ is the highest frequency that includes information and $T$ is the duration of the gravitational wave signal in the sensitive frequency band of the detector. The computational cost can be especially prohibitive if $T$ is long or if $f_{\rm max}$ is large. Even non-uniform frequency sampling still requires a large number of grid points, as the waveform phase typically evolves over a large number of cycles across the sensitive frequency band. For GW parameter inference with sophisticated waveform models, model calls at individual frequencies often dominate the runtime. For many frequency domain waveform models, the computational time of a likelihood evaluation scales nearly linearly with the number of queried frequencies \cite{roq}.

A number of methods have been developed to approximate the likelihood with fewer frequency evaluations, trading an insignificantly amount of accuracy degrade for substantial speedup. Reduced order quadrature (ROQ) \cite{roq} is one such method, which pre-computes a reduced basis for a given waveform model, and approximates any physical waveform with an interpolant in the linear space spanned by this basis. This method only needs as many independent evaluations as there are basis elements, and has been applied to reduce the computational time of a variety of waveform models \cite{roqPhenomPv2, roqSEOBNRv2, gen3bayesian, focusedroq}.

Relative binning~\cite{relativebinning, relativebinning170817, extendedrelativebinning}, originally developed for non-precessing waveforms in the dominant $(2,\,2)$ radiation mode, is another method to achieve the same goal. In relative binning, the {\it ratio} between a given waveform and an appropriately chosen fiducial waveform is considered.  If the fiducial waveform is a decent fit of the data, only waveforms that resemble the fiducial one need to be examined in parameter inference, for which the waveform ratio conveniently has a milder frequency dependence. For this reason, the waveform ratio can be well approximated as a piecewise-linear function in the frequency domain, which only requires waveform model calls at the edges of a reduced number of frequency bins. 

For simple waveform models with only one harmonic such as \texttt{IMRPhenomD} \cite{PhenomD}, this method has been shown to require fewer waveform evaluations than ROQ for the same error tolerance~\cite{relativebinning}, and has been routinely used in independent analysis of public LIGO/Virgo data \cite{GW151226Zackay, IASo2Venumadhav, Roulet_2020, roulet2021distribution, nitz20213ogc}. An earlier and somewhat similar method referred to as heterodyne, which also samples the ratio between two waveforms using fewer grid points, was proposed in Ref.~\cite{lazylikelihoods}.

Many gravitational wave sources carry a strong imprint of the effects of precession of the orbital plane, which are neglected in simpler models like \texttt{IMRPhenomD}. These precession effects are often accounted for by effective precession parameters that well approximate the effects with minimal additional complexity or dimensionality \cite{ogchip, newchip, effectivespinvector}. However, more sophisticated waveform models that fully incorporate the effects of spin-orbit precession with misaligned spins and higher harmonic modes~\cite{PhenomP, IMRPhenomPv3HM, twistagain, SEOBNRv4PHM, NRSur7dq4} have begun to be widely applied in parameter inference. Compared to many of the non-precessing models, which only incorporate the dominant $(2,\,2)$ radiation mode, these advanced models have yielded better fits to numerical simulations \cite{Garcia-Quiros:2020qpx}, better fits to the detected LIGO/Virgo gravitational wave signals \cite{GW170729HM}, higher SNR, more precision, and more accuracy in parameter estimation for injected signals \cite{PhenomHMPE, NRHybSur3dq8PE, EOBHMPE}, and in some cases have led to the intriguing discovery of new solutions~\cite{addinghm, GW190814, GW151226alternate, GW190521alternate}. 

Additionally, information in higher modes is crucial in tracking the recoil kicks of the GW sources \cite{LinearKicks}, and this information should be reliably extractable once LIGO/Virgo reach their design sensitivities \cite{LinearKicksTest}.

In the original implementation of relative binning presented in Ref. \cite{relativebinning}, the key assumption is that the ratio between a waveform under examination over a fiducial one has significantly smoother frequency dependence than the waveform themselves do. The quality of this assumption however degrades for precessing waveforms with the full symphony of harmonics, as both the precession-induced wobbling of the binary orbital plane and the superposition of multiple harmonics give rise to additional oscillations with the frequency in the observed waveform. It is pivotal to realize that a precessing waveform can be decomposed into a linear combination of contributions from individual radiation harmonics in the co-precessing frame \cite{Schmidt:2010it}, and our schemes deal with ratios evaluated for these modes individually and hence only have to sample functions that are smoother in the frequency space. Our scheme builds a likelihood from piecewise linear interpolants of these smooth quantities. As an explicit example, we demonstrate this strategy using the phenomenological waveform model \texttt{IMRPhenomXPHM} \cite{twistagain,Garcia-Quiros:2020qpx,Pratten:2020fqn}.

In this paper, we present two improved relative binning schemes specially designed for precessing waveforms with multiple harmonics, which we jointly refer to as the Mode-by-mode Relative Binning (MRB). We give explicit formulae to approximate the likelihood, and we introduce practical algorithms to choose appropriate frequency bins. We find theoretical speedup of MRB relative to the original relative binning method by up to an order of magnitude, which is due to a reduced number of frequency bins required, for the same tolerable {\it absolute} accuracy $\sim 0.1$ in the log likelihood. 

The rest of the paper is organized as follows. In \refsec{twistingprocedure}, we review the ``twisting-up'' procedure for building precession waveforms and introduce the basic definitions that will be used in the MRB schemes. In \refsec{modebymoderelativebinning}, we present the technical details of the two MRB schemes, including introducing piecewise linear approximation for quantities that are smooth functions of the frequency, as well as new methods to optimize the frequency bins. In \refsec{testingmethods}, we study the required number of frequency bins as a function of accuracy tolerance by applying MRB to real and injected gravitational wave signals, with a comparison to the old relative binning method without using mode decomposition. Finally, we provide discussion of our results in \refsec{discussion} and concluding remarks in \refsec{concl}.

\section{Mode decomposition for precessing waveforms}
\label{sec:twistingprocedure}

To understand how relative binning can be best applied to waveforms that describe precessing binary systems, we start by discussing the general structure of precessing waveforms. Many precessing models with higher modes are built using an approximate mapping from a non-precessing system to a precessing system, often referred to as the ``twisting-up'' procedure \cite{Schmidt:2010it,twistingup}. This includes state-of-the-art phenomenological \cite{IMRPhenomPv3HM,twistagain} and effective one body \cite{SEOBNRv4PHM} waveform models. Under the twisting-up scheme, the non-precessing waveforms resemble the gravitational radiation emitted by a non-precessing source, as if an imaginary observer is fixed relative to the wobbling orbital plane of the binary \cite{Schmidt:2010it}. 

The specific prescription for twisting-up the non-precessing waveforms that we use is developed in \cite{twistingup} and is fully detailed for the \texttt{IMRPhenomXPHM} model in \cite{twistagain}. Here, we note the most important aspects of this prescription that will allow us to explain our schemes. A general waveform model outputs the frequency domain waveform of both polarizations, i.e. $h_+(f)$ and $h_\times(f)$. Models that use the twisting-up procedure include a prescription for a non-precessing waveform as what a fictitious observer would see in the co-precessing frame. This non-precessing waveform can be decomposed into building blocks, $h^L_{\ell, m'}(f)$, following the spin-weighted spherical harmonic expansion. Since the binary orbital timescale is typically significantly shorter than the timescale of spin-orbit precession, the precession effects on the waveform can be well captured by a time-dependent Euler rotation of the non-precessing waveform \cite{twistingup}. After transforming from the time domain into the frequency domain under the stationary phase approximation, the model also provides three frequency-dependent Euler angles, $\alpha(f)$, $\beta(f)$ and $\gamma(f)$, which parameterize a rotation from the frame co-precessing with the orbital angular momentum vector (the $L$-frame) to the approximately inertial frame defined by the total angular momentum vector (the $J$-frame). Using the definitions introduced in Ref.~\cite{twistagain}, it is a straightforward derivation that both polarizations can be expressed as a linear combination of the $L$-frame modes $h^L_{\ell, m'}$. We follow their convention that $m$ is used to label the $J$-frame modes, and $m'$ is used to label the $L$-frame modes.
\begin{subequations}
\begin{equation}
    h_+(f) = \sum_{\ell, m'}\,C^+_{\ell, m'}(f)\,h^L_{\ell, m'}(f),
\label{eq:hplus}
\end{equation}
\begin{equation}
    h_-(f) = \sum_{\ell, m'}\,C^\times_{\ell, m'}(f)\,h^L_{\ell, m'}(f).
\label{eq:hcross}
\end{equation}
\label{eq:hpluscross}
\end{subequations}

Often, only a few subdominant modes are included in this sum as the contribution of almost all higher order modes are negligibly small. The dominant $(\ell,\,m')=(2,\,2)$ mode is always included, and the leading subdominant modes $(\ell,\,m')=(2,\,1)$ and $(\ell,\,m')=(3,\,3)$ are usually included. Our implementation and the \texttt{LALSuite} implementation \cite{lalsuite} of the \texttt{IMRPhenomXPHM} model use these modes and the $(\ell,\,m')=(3,\,2)$ and $(\ell,\,m')=(4,\,4)$ modes. Hence, the overall waveform observed in the inertial frame is reduced to a short list of basis modes, which have simpler frequency dependence individually.

The frequency-dependent coefficients $C^+_{\ell, m'}(f)$ and $C^\times_{\ell, m'}(f)$ incorporate an in-plane rotation by a frequency-independent angle $\zeta$, which depends on the orientation of the binary, to correct for the difference between the computed spin-weighted spherical harmonics defined with respect to the total angular momentum vector, and the conventional parameterization of the waveform defined with respect to the reference orbital angular momentum vector.

\begin{subequations}
\begin{equation}
    C^+_{\ell, m'}(f) = \cos(2\zeta)\,\tilde{C}^+_{\ell, m'}(f) + \sin(2\zeta)\,\tilde{C}^\times_{\ell, m'}(f),
\label{eq:Cplus}
\end{equation}
\begin{equation}
    C^\times_{\ell, m'}(f) = \cos(2\zeta)\,\tilde{C}^\times_{\ell, m'}(f) - \sin(2\zeta)\,\tilde{C}^+_{\ell, m'}(f).
\label{eq:Ccross}
\end{equation}
\label{eq:Cpluscross}
\end{subequations}
The tilded coefficients are specified in terms of the frequency-dependent Euler angles $\alpha(f)$, $\beta(f)$ and $\gamma(f)$, as well as $\theta_{JN}$, the constant angle between the total angular momentum vector and the line of sight:
\begin{subequations}
\begin{equation}
    \tilde{C}^+_{\ell, m'}(f) = \frac{1}{2}\,e^{im'\gamma(f)}\sum_{m}\left(A^\ell_{m,-m'}(f)+(-1)^\ell \left[A^\ell_{m,m'}(f)\right]^*\right),
\label{eq:Ctildeplus}
\end{equation}
\begin{equation}
    \tilde{C}^\times_{\ell, m'}(f) = \frac{i}{2}\,e^{im'\gamma(f)}\sum_{m}\left(A^\ell_{m,-m'}(f)-(-1)^\ell \left[A^\ell_{m,m'}(f)\right]^*\right),
\label{eq:Ctildecross}
\end{equation}
\label{eq:Ctildepluscross}
\end{subequations}
where $^*$ stands for complex conjugation, and we introduce the following coefficient to describe the mixing between the $L$-frame modes and the $J$-frame modes under precession:
\begin{equation}
    A^\ell_{m,m'}(f)=e^{-im\alpha(f)}\,d^\ell_{m,m'}(\beta(f))_{-2}\,Y_{\ell,m}(\theta_{JN},0).
\label{eq:Atransfer}
\end{equation}
Here, $d^\ell_{m,m'}(\beta(f))$ is the Wigner small $d$-matrix element evaluated at the Euler angle $\beta(f)$, and $_{-2}Y_{\ell,m}$ is the spin-weighted spherical harmonic.

Finally, the observed strain $h(f)$ at a given detector is a linear combination of the two wave polarizations $h_+(f)$ and $h_\times(f)$. The linear coefficients are given by the detector response coefficients, $F_+$ and $F_\times$, which are dependent on the sky position of the gravitational wave source and the orientation of the detector at the time of the gravitational wave event. The detector response coefficients depend on three extrinsic parameters: right ascension $\alpha$, declination $\delta$, and the polarization angle $\psi$. In the frequency domain, the arrival time $t_0$ of the event introduces a phase that scales linearly with the frequency.
\begin{equation}
    h(f) = \left( F_+(\alpha, \delta, \psi)\,h_+(f) + F_\times(\alpha, \delta, \psi)\, h_\times(f)\right)\,e^{-2\pi i f t_0}.
\label{eq:strain}
\end{equation}
The overall result is that the strain observed at a given detector can be cast into the following linear combination of the non-precessing modes,
\begin{equation}
    h(f) = \sum_{\ell, m'}\,C_{\ell, m'}(f)\,h^L_{\ell, m'}(f)\,e^{-2\pi i f t_0},
\label{eq:modedecomp}
\end{equation}
where
\begin{equation}
    C_{\ell, m'}(f) = F_+(\alpha, \delta, \psi)\,C^+_{\ell,m'}(f) + F_\times(\alpha, \delta, \psi)\,C^\times_{\ell,m'}(f).
\label{eq:Ccoeffs}
\end{equation}
Two additional definitions are convenient for discussion in the following sections. First, we can absorb the arrival time dependence into the $L$-frame mode to define the time-dependent quantity
\begin{equation}
    \hat{h}_{\ell, m'}(f) = h^L_{\ell, m'}(f)\,e^{-2\pi i f t_0},
\label{eq:timedependentLframemode}
\end{equation}
for which we simply have
\begin{equation}
    h(f) = \sum_{\ell, m'} C_{\ell, m'}(f)\, \hat{h}_{\ell, m'}(f),
\label{eq:timedependentLframemodedecomp}
\end{equation}
Second, we can incorporate the coefficient $C_{\ell, m'}$ and define the full mode component
\begin{equation}
    h_{\ell, m'}(f) = C_{\ell, m'}(f)\, h^L_{\ell, m'}(f)\, e^{-2\pi i f t_0}.
\label{eq:modecomponent}
\end{equation}
The observed waveform then has the simple decomposition
\begin{equation}
    h(f) = \sum_{\ell, m'} h_{\ell, m'}(f).
\label{eq:modecomponentmodedecomp}
\end{equation}

\section{Mode-by-mode Relative Binning Schemes}
\label{sec:modebymoderelativebinning}

As mentioned in \refsec{intro}, the original relative binning scheme assumes that the full waveform divided by a fiducial waveform is smooth in frequency space so that it can be well approximated by a piecewise linear interpolant. This scheme was developed for non-precessing waveform models that only account for the dominant $(2,\,2)$ mode. When applied to these waveform models, the assumption of the original relative binning scheme is only that the ratio of the $(2,\,2)$ mode evaluated for any given parameters over the $(2,\,2)$ mode evaluated at a fiducial set of parameters is a smooth function of the frequency. If the original relative binning method is applied to a precessing waveform comprised of multiple modes, this assumption is being extended to the sum of all modes, which is a stronger assumption. Instead, we improve our accuracy by not extending this assumption and only applying it to individual {\it non-precessing} modes. 

\begin{figure}[!]
\begin{tabular}{c}
\includegraphics[width=85mm]{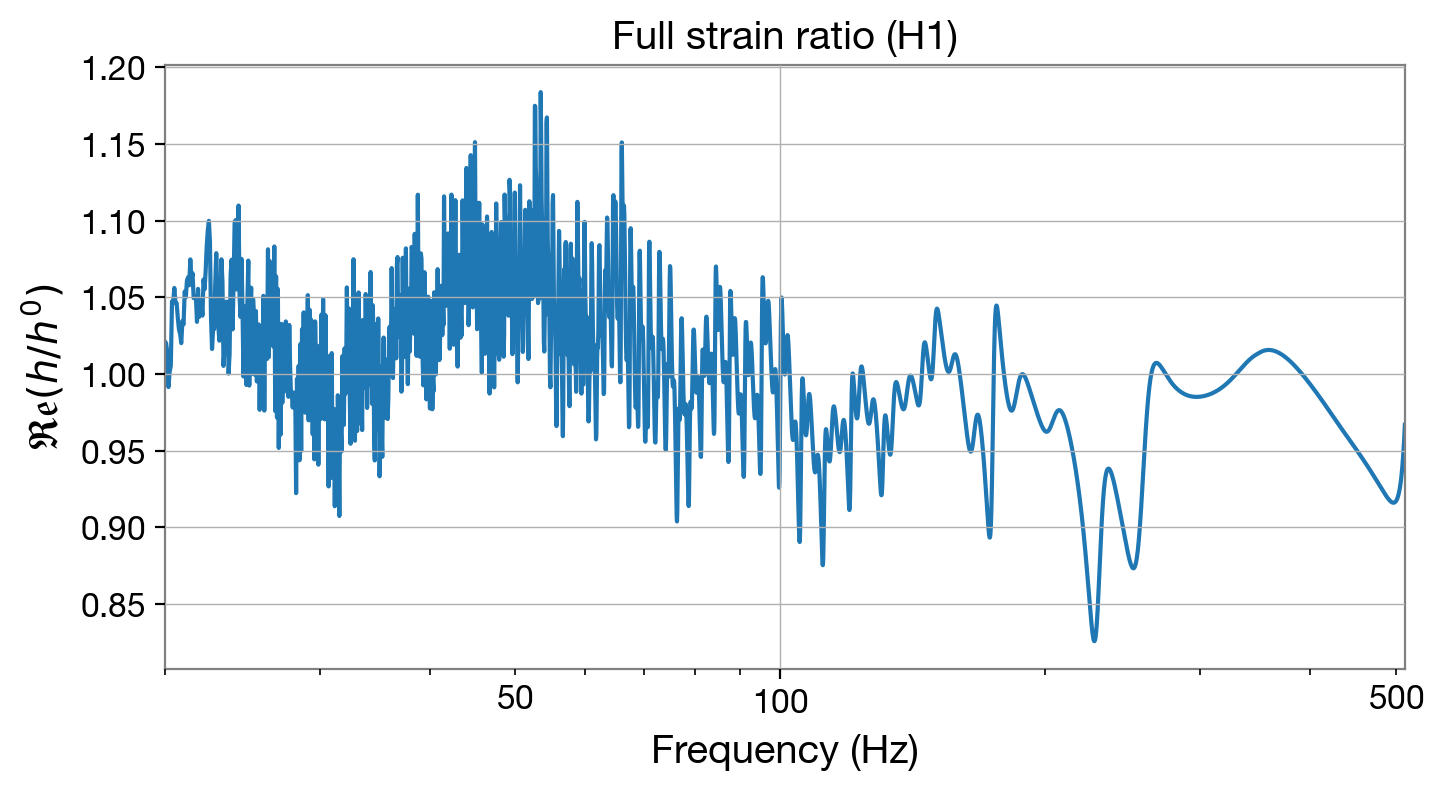} \\
\includegraphics[width=85mm]{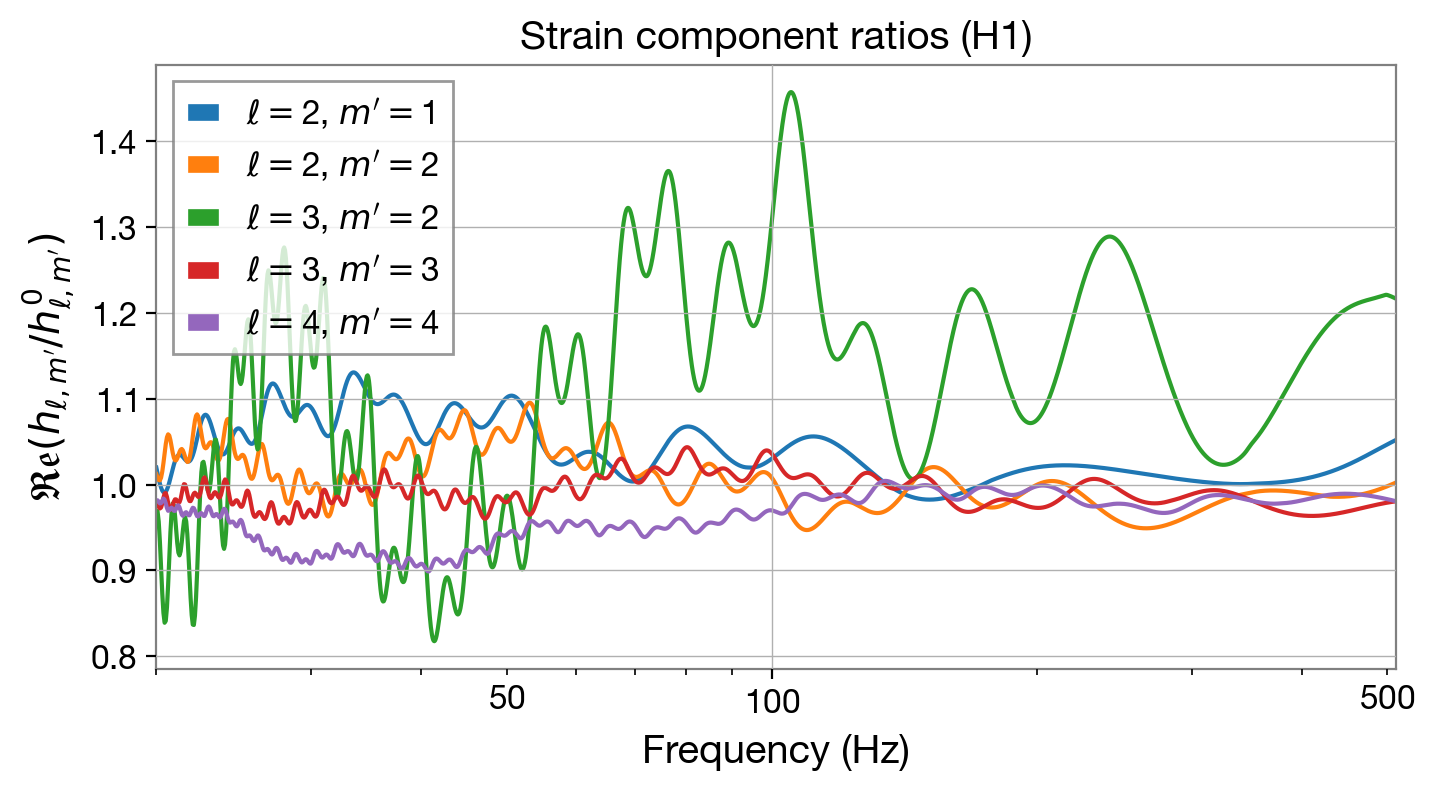} \\
\includegraphics[width=85mm]{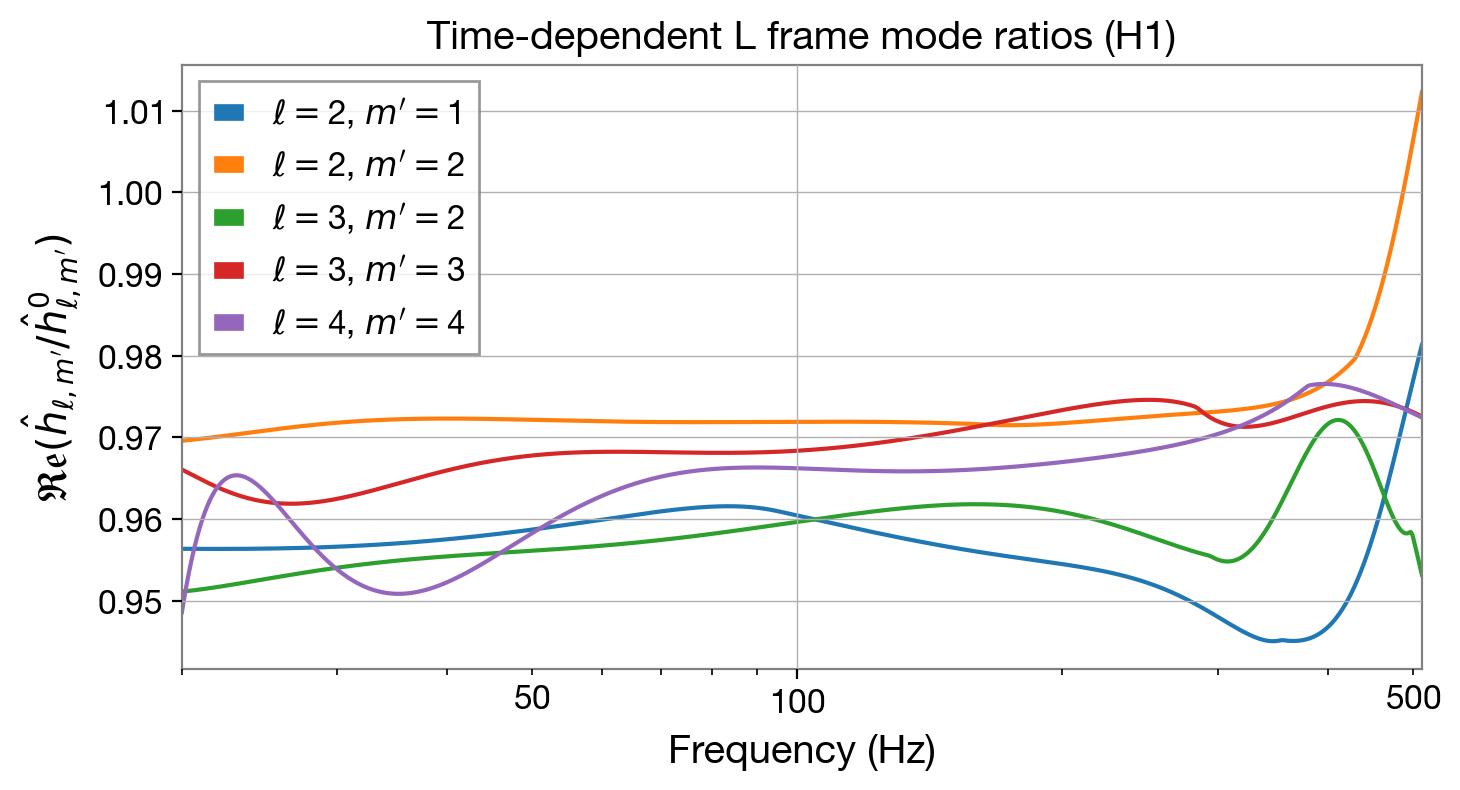} \\
\includegraphics[width=85mm]{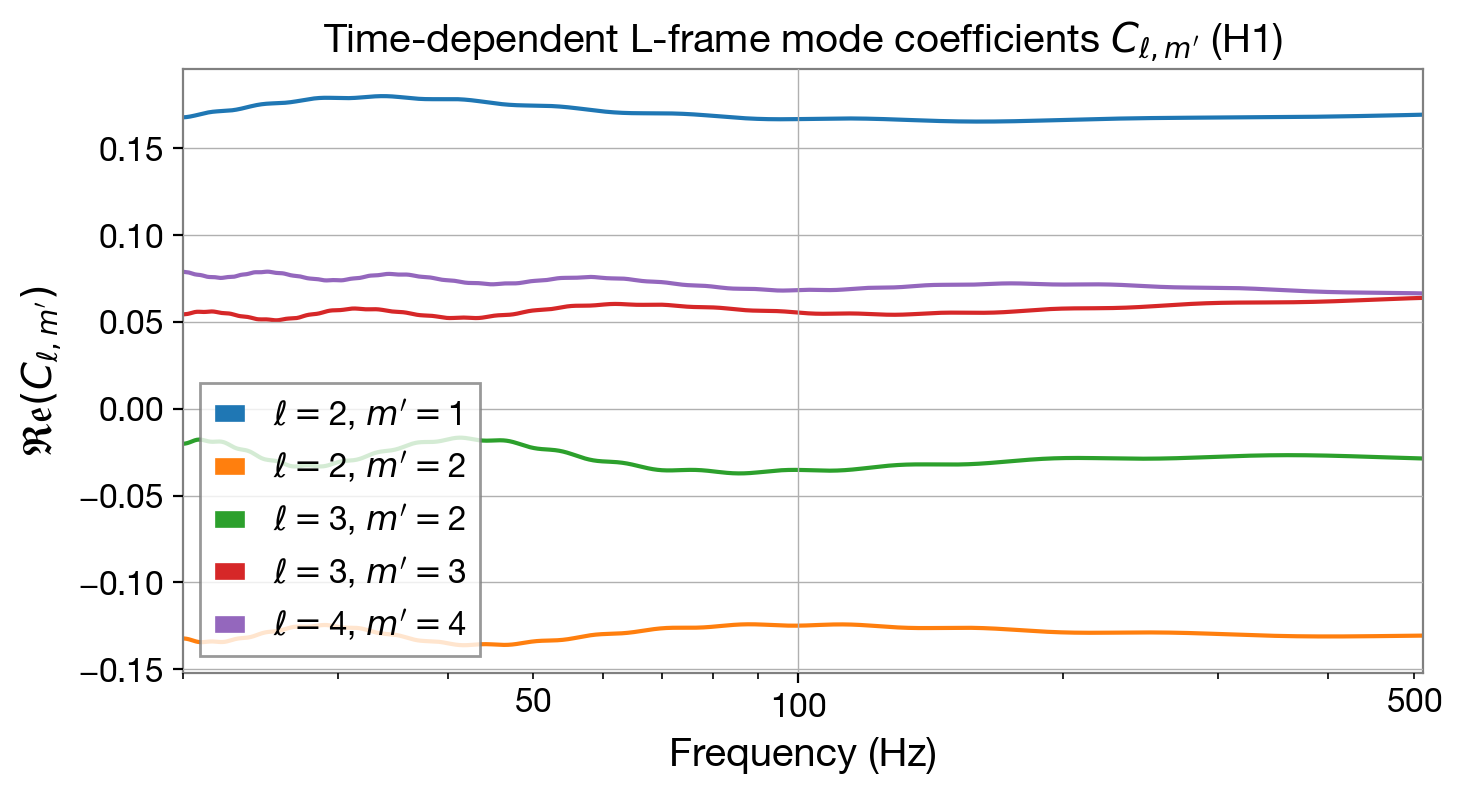}
\end{tabular}
\caption{Functional smoothness as a function of frequency for the components or ratios that are approximated by piecewise linear functions in different relative binning methods. Shown are the real part of these complex-valued quantities, evaluated at the LIGO Hanford detector as an example. The fiducial waveform and one sample waveform are chosen from the posterior samples collected from our \texttt{MultiNest} run on the gravitational wave event GW190814. The harmonics $(\ell,\,m)$ shown are the 5 modes that are included by default in the the waveform model \texttt{IMRPhenomXPHM}. The original relative binning method linearizes the full waveform, as shown in the top panel. Mode-by-mode relative binning linearizes the time-dependent $L$-frame mode ratios and their coefficients (bottom two panels) under Scheme 1, and the strain component ratios (second panel) under Scheme 2.}
\label{fig:smooth}
\end{figure}

At a fixed orbital frequency, the various $L$-frame modes produced by a binary source have different frequencies, and hence their superposition exhibit interference features. This leads to a jagged ratio for the overall waveform, $h(f)$, as a function of the frequency, as we exemplify in the top panel of \reffig{smooth}. The lower three panels in \reffig{smooth}, by contrast, show the smoother quantities $h_{\ell, m'}$, defined in \refeq{modecomponent}, $\hat h_{\ell, m'}$, defined in \refeq{timedependentLframemode}, and $C_{\ell, m'}$, defined in \refeq{Ccoeffs}, which are associated with individual $L$-frame modes. We will instead approximate these quantities or their ratios over fiducial counterparts by piecewise linear interpolants in our two schemes.

\subsection{Scheme 1}

In the first scheme, we handle each time-dependent $L$-frame mode $\hat h_{\ell, m'}(f)$ as we handle the full waveform in the original relative binning method. To this end, we decompose the fiducial waveform as
\begin{equation}
    h^0(f) = \sum_{\ell, m'}\,C^0_{\ell, m'}(f)\,\hat{h}^0_{\ell, m'}(f).
\label{eq:fiducialmodedecomp1}
\end{equation}
In this scheme, we suitably create frequency bins which we label by $b$. Within each frequency bin, the ratio of the time-dependent $L$-frame mode over some fiducial one is well approximated by a linear interpolant:
\begin{equation}
    r_{\ell, m'}(f) = \frac{\hat{h}_{\ell,m'}(f)}{\hat{h}^0_{\ell,m'}(f)}\approx r^0_{\ell,m'}(h, b) + r^1_{\ell, m'}(h,b)\,\left(f-f_c(b)\right),
\label{eq:ratio1}
\end{equation}
where $f_c(b)$ is the frequency at the bin center. The coefficients $r^0_{\ell,m'}(h, b)$ and $r^1_{\ell, m'}(h,b)$, defined for every frequency bin, are found by computing the waveform ratio at the bin edges. The speedup of relative binning hence comes from the fact that it suffices to evaluate at the bin edges for an excellent approximation of the likelihood. To obtain the full waveform $h(f)$, we also need the frequency dependent coefficients, which we also linearize within each frequency bin:
\begin{equation}
    C_{\ell, m'}(f) \approx C^0_{\ell,m'}(h, b) + C^1_{\ell, m'}(h,b)\left(f-f_c(b)\right).
\label{eq:linearC}
\end{equation}
The constant coefficient in the piecewise linear interpolant of $C$ ($\,C^0_{\ell,m'}(h, b)\,$) is not to be confused with the fiducial $C$ ($\,C^0_{\ell, m'}(f)\,$). We disambiguate by showing that the former is a function of bin $b$ and waveform $h$, and the latter is a function of frequency $f$. Likewise, the coefficients $C^0_{\ell,m'}(h, b)$ and $C^1_{\ell,m'}(h, b)$ can be computed from $C_{\ell, m'}(f)$ evaluated at the bin edges. In practice, we find that both quantities are smooth functions of the frequency, as shown in \reffig{smooth}. We then need to compute the likelihood in terms of these linearized ratios and coefficients. The log likelihood for the data series, $d(f)$, given a signal series $h(f)$, is given by:
\begin{equation}
    \ln L = \mathfrak{Re}\,Z[d,h] - \frac{1}{2}Z[h,h],
\label{eq:lnL}
\end{equation}
where $Z[\cdots]$ is the notation for the standard complex-valued overlaps defined via a summation over only positive frequencies $f>0$:
\begin{subequations}
\begin{equation}
    Z[d,h] = 4\,\sum_f \frac{d(f)\,h^*(f)}{S_n(f)/T},
\label{eq:Zdh}
\end{equation}
\begin{equation}
    Z[h,h] = 4\,\sum_f \frac{|h(f)|^2}{S_n(f)/T},
\label{eq:Zhh}
\end{equation}
\label{eq:overlaps}
\end{subequations}
where $S_n(f)$ is one-sided power spectral density (PSD) of the detector noise, and $T$ is the length of time series analyzed.

To approximate the above overlaps, we introduce a number of summary data for each frequency bin. The $A$ coefficients are used in the overlap $Z[d,h]$ and are associated with a single $L$-frame mode because they are linear in the strain $h$. The $B$ coefficients are used in the overlap $Z[h,h]$ and are associated with a pair of $L$-frame modes because they are quadratic in the strain $h$.
\begin{widetext}
\begin{subequations}
\begin{equation}
    A^0_{\ell, m'}(b) = 4\,\sum_{f\in b} \frac{d(f)\,\hat{h}^{0*}_{\ell, m'}(f)}{S_n(f)/T}
\label{eq:A01}
\end{equation}
\quad
\begin{equation}
    A^1_{\ell, m'}(b) = 4\,\sum_{f\in b} \frac{d(f)\,\hat{h}^{0*}_{\ell, m'}(f)}{S_n(f)/T}\left(f-f_c(b)\right)
\label{eq:A11}
\end{equation}
\begin{equation}
    B^0_{\ell, m', \tilde{\ell}, \tilde{m}'}(b) = 4\,\sum_{f\in b} \frac{\hat{h}^{0}_{\ell, m'}(f)\,\hat{h}^{0*}_{\tilde{\ell}, \tilde{m}'}(f)}{S_n(f)/T}
\label{eq:B01}
\end{equation}
\quad
\begin{equation}
    B^1_{\ell, m', \tilde{\ell}, \tilde{m}'}(b) = 4\,\sum_{f\in b} \frac{\hat{h}^{0}_{\ell, m'}(f)\,\hat{h}^{0*}_{\tilde{\ell}, \tilde{m}'}(f)}{S_n(f)/T}\left(f-f_c(b)\right)
\label{eq:B11}
\end{equation}
\label{eq:summarydata1}
\end{subequations}
\end{widetext}
These summary data only depend on the data series $d(f)$ and the chosen fiducial waveform, and hence can be computed in advance of the large number of likelihood evaluations required in the sampling process. To obtain the log likelihood at each sampled set of waveform parameters, we approximate the overlaps up to linear order in $(f-f_c(b))$, using the following sums over frequency bins
\begin{widetext}
\begin{subequations}
\begin{align}
    &Z[d,h]=\sum_b\sum_{\ell, m'} \left[ A^0_{\ell, m'}(b)\,r^{0*}_{\ell,m'}(h, b)\,C^{0*}_{\ell,m'}(h, b) + A^1_{\ell, m'}(b)\,\left(r^{1*}_{\ell,m'}(h, b)\,C^{0*}_{\ell,m'}(h, b) + r^{0*}_{\ell,m'}(h, b)\,C^{1*}_{\ell,m'}(h, b) \right) \right],
\label{eq:binningZdh1}
\end{align}
\begin{align}
    Z[h,h]&=\sum_b\sum_{\ell, m'}\sum_{\tilde{\ell}, \tilde{m}'} \left\{ B^0_{\ell, m', \tilde{\ell}, \tilde{m}'}(b)\,r^{0}_{\ell,m'}(h, b)\,r^{0*}_{\tilde{\ell},\tilde{m}'}(h, b)\,C^{0}_{\ell,m'}(h, b)\,C^{0*}_{\tilde{\ell}, \tilde{m}'}(h, b) \right. \nonumber \\
    & \left. + B^1_{\ell, m', \tilde{\ell}, \tilde{m}'}(b)\,\left[C^{0}_{\ell,m'}(h, b)\,C^{0*}_{\tilde{\ell}, \tilde{m}'}(h, b)\,\left(r^{0}_{\ell,m'}(h, b)\,r^{1*}_{\tilde{\ell},\tilde{m}'}(h, b) + r^{0*}_{\tilde{\ell},\tilde{m}'}(h, b)\,r^{1}_{\ell,m'}(h,b)\right) \right.\right.\nonumber \\
    & \left.\left. + r^{0}_{\ell,m'}(h, b)\,r^{0*}_{\tilde{\ell},\tilde{m}'}(h, b)\,\left( C^{0}_{\ell,m'}(h, b)\,C^{1*}_{\tilde{\ell},\tilde{m}'}(h, b) + C^{0*}_{\tilde{\ell},\tilde{m}'}(h, b)\,C^{1}_{\ell,m'}(h,b)\right) \right] \right\}.
\label{eq:binningZhh1}
\end{align}
\label{eq:binningZ1}
\end{subequations}
\end{widetext}
To summarize, the summary data \refeq{summarydata1} can be pre-computed from the data and the chosen fiducial waveform; then to approximate the log likelihood using \refeq{binningZ1} and \refeq{lnL}, $r^0_{\ell,m'}(h, b)$, $r^1_{\ell, m'}(h,b)$, $C^0_{\ell,m'}(h, b)$, and $C^1_{\ell,m'}(h, b)$ only need to be evaluated at the bin edges.

\subsection{Scheme 2}

Scheme 1 involves the multiplication of two quantities, both of which are separately approximated as piecewise linear functions. One might wonder if this strategy accumulates excessive errors. Hence, we consider an alternative scheme, which we refer to as Scheme 2. In this scheme, we linearize the ratio of the entire mode component over its fiducial counterpart. To do so, we decompose the fiducial waveform $h^0(f)$ as in \refeq{modecomponentmodedecomp}:
\begin{equation}
    h^0(f) = \sum_{\ell, m'} h^0_{\ell, m'}(f)
\label{eq:fiducialmodedecomp2}
\end{equation}
Within the frequency bin $b$, we make the approximation
\begin{equation}
    r_{\ell, m'}(f) = \frac{h_{\ell,m'}(f)}{h^0_{\ell,m'}(f)}\approx r^0_{\ell,m'}(h, b) + r^1_{\ell, m'}(h,b)\left(f-f_c(b)\right).
\label{eq:ratio2}
\end{equation}
This quantity is also much smoother than the ratio for the observed waveform, which is linearized in the original relative binning method, as one can see in \reffig{smooth}. In this scheme, our summary data are very similar, but they use the full fiducial mode components:
\begin{subequations}
\begin{equation}
    A^0_{\ell, m'}(b) = 4\sum_{f\in b} \frac{d(f)\,h^{0*}_{\ell, m'}(f)}{S_n(f)/T}
\label{eq:A02}
\end{equation}
\begin{equation}
    A^1_{\ell, m'}(b) = 4\sum_{f\in b} \frac{d(f)\,h^{0*}_{\ell, m'}(f)}{S_n(f)/T}\left(f-f_c(b)\right)
\label{eq:A12}
\end{equation}
\begin{equation}
    B^0_{\ell, m', \tilde{\ell}, \tilde{m}'}(b) = 4\sum_{f\in b} \frac{h^{0}_{\ell, m'}(f)\,h^{0*}_{\tilde{\ell}, \tilde{m}'}(f)}{S_n(f)/T}
\label{eq:B02}
\end{equation}
\begin{equation}
    B^1_{\ell, m', \tilde{\ell}, \tilde{m}'}(b) = 4\sum_{f\in b} \frac{h^{0}_{\ell, m'}(f)\,h^{0*}_{\tilde{\ell}, \tilde{m}'}(f)}{S_n(f)/T}\left(f-f_c(b)\right)
\label{eq:B12}
\end{equation}
\label{eq:summarydata2}
\end{subequations}
Now that there is only one linearized quantity, the overlaps have shorter expressions:
\begin{subequations}
\begin{equation}
    Z[d,h]=\sum_b\sum_{\ell, m'} \left[ A^0_{\ell, m'}(b)\,r^{0*}_{\ell,m'}(h, b) + A^1_{\ell, m'}(b)\,r^{1*}_{\ell,m'}(h, b) \right],
\label{eq:binningZdh2}
\end{equation}
\begin{align}
    &Z[h,h]=\sum_b\sum_{\ell, m'}\sum_{\tilde{\ell}, \tilde{m}'} \left[  B^0_{\ell, m', \tilde{\ell}, \tilde{m}'}(b)\,r^{0}_{\ell,m'}(h, b)\,r^{0*}_{\tilde{\ell},\tilde{m}'}(h, b) \right. \nonumber \\
    & \hspace{-1cm} \left. + B^1_{\ell, m', \tilde{ell}, \tilde{m}'}(b)\left(r^{0}_{\ell,m'}(h, b)\,r^{1*}_{\tilde{\ell},\tilde{m}'}(h, b)+r^{0*}_{\tilde{\ell},\tilde{m}'}(h, b)\,r^{1}_{\ell,m'}(h,b)\right) \right].
\label{eq:binningZhh2}
\end{align}
\label{eq:binningZ2}
\end{subequations}
Again, the pre-computed summary data \refeq{summarydata2} can be substituted into \refeq{binningZ2} to compute the overlaps \refeq{overlaps} up to linear order in $(f-f_c(b))$. The coefficients $r^0_{\ell,m'}(h, b)$ and $r^1_{\ell, m'}(h,b)$ only require evaluating the model waveform at the edges of the frequency bins. Although Scheme 2 have simpler expressions than Scheme 1, it is not meaningfully faster in terms of the computation runtime, as all evaluations needed for computing $C_{\ell, m'}(f)$ and $\hat{h}_{\ell, m'}(f)$ are also needed for computing $h_{\ell, m'}(f)$.

\subsection{Bin Selection}

Previously, a prescription to determine the frequency bins for the original relative binning method is presented in Ref.~\cite{relativebinning}. Motivated by the post-Newtonian expansion, this prescription selects frequency bins to bound the differential phase changes assuming a phenomenological ansatz in the form of a sum over power-law phases. Although the prescription has the advantage that it is independent of the specific compact binary coalescence signals under analysis, it may not be the optimized bin selection on the event-by-event basis. In particular, for high binary masses, the merger and ringdown signals may dominate the sensitive frequency band of the detector, and the corresponding gravitational wave signal may be poorly described by the post-Newtonian expansion. 

In this work, we present a new bin selection algorithm that is adaptive to any given gravitational wave signal under investigation. Since the linearized quantities associated with individual modes have smoother frequency dependence than the observed waveform, we can afford fewer frequency bins to achieve a comparable accuracy for the log likelihood. Since the bottleneck in the computational time cost is still the large number of likelihood evaluations, we can afford to adjust the frequency bins event by event via an algorithm that is more time-consuming than the one that is adopted in Ref.~\cite{relativebinning}. We choose a representative waveform (different from the fiducial waveform) for an estimate of the binning errors. We can then test out various binning choices on this representative waveform. Methods for choosing this waveform are discussed in \refsec{discussion}.

Our algorithm determines the frequency bins by starting with the entire frequency range as a single bin and iteratively bisecting the frequency bins. In this iterative process, we compute, for each existing frequency bin, the error contribution to the log likelihood due to mode-by-mode relative binning. We keep bisecting the bins until the overall absolute error in the log likelihood is smaller than a empirically set threshold. We iterate the whole procedure and update the target number of bins to the number of bins achieved in the previous run, until the resulting number of bins and the target number of bins converge. This algorithm is detailed in \refalg{binselection}.

\begin{figure}
\begin{algorithm}[H]
   \caption{Bin Selection}
    \begin{algorithmic}[1]
    \Function{getBins}{$\eta$, targetN=200, testParams}
        \State targetBinError = $\eta$/targetN
        \State candidateBins = \Call{bisectBinSearch}{targetBinError, testParams, fullFrequencyGrid}
        \State nBins = \Call{numberOfBins}{candidateBins}
        \If {nBins = targetN}
            \State \Return candidateBins
        \Else
            \State \Return \Call{getBins}{$\eta$, targetN=nBins, testParams}
        \EndIf
    \EndFunction
    \bigskip
    \Function{bisectBinSearch}{targetBinError, testParams, candidateBin}
        \If {\Call{logLikelihoodError}{testParams, candidateBin} $\leq$ targetBinError} 
            \State \Return candidateBin
        \ElsIf{candidateBin is minimum size}
            \State \Return candidateBin
        \Else
            \State \Return \Call{bisectBinSearch}{targetBinError, testParams, leftHalf} +  \Call{bisectBinSearch}{targetBinError, testParams, rightHalf}
        \EndIf
    \EndFunction
\end{algorithmic}
\label{alg:binselection}
\end{algorithm}
\end{figure}
The algorithm consistently converges to a fixed number of bins independent of the initial target number of bins (targetN), so we set this number arbitrarily to 200. By varying the total allowed error $\eta$ on the test parameters, we trade computational cost (number of bins) for accuracy. We tested similar algorithms that unevenly divide the bins, but found that such methods are slower and did not improve accuracy.

\section{Testing Methods and Results}
\label{sec:testingmethods}

We tested the accuracy and efficiency of both mode-by-mode relative binning schemes on several real and synthetic compact binary coalescence events. We computed likelihoods of the \texttt{IMRPhenomXPHM} model using \texttt{LALSuite} \cite{lalsuite} and gathered posterior samples using the \texttt{MultiNest} sampler \cite{multinest}.

To test our method, we consider gravitational wave signals with either strong spin-orbit precession or large contributions from the subdominant modes, or a combination of both, as their parameter estimation can be greatly improved with \texttt{IMRPhenomXPHM} or other precessing models with higher modes that exploit the twisting-up procedure. Precession occurs most dramatically in binaries that have large spin vectors misaligned with the binary orbital angular momentum vector (i.e. large in-plane spin components) \cite{precession}. Odd parity subdominant modes such as $(3,\,3)$ are important when there is a large asymmetry between the two components, e.g. when the mass ratio $q = m_2/m_1 < 1$ is small. Also, subdominant modes are larger in the late stage of the merger. For low mass mergers, the most sensitive LIGO/Virgo frequency band corresponds to the early inspiral of the merger. On the contrary, for heavy mass mergers, the most sensitive frequency band corresponds to the merger stage when subdominant modes are stronger. For these reasons, we study events with large in-plane spin components, or a small mass ratio $q$, or a large total mass.

We consider one real LIGO/Virgo event, GW190814, and four injected events. GW190814 is chosen because the original analysis shows strong Bayesian evidence favoring a model that includes the $(3,\,3)$ mode over one that has only the dominant $(2,\,2)$ mode \cite{GW190814}. The four injections are chosen as the following. Two are chosen to resemble the inferred source properties of two interesting LIGO/Virgo detections for which new source solutions have been recently reported from independent reanalyses using the \texttt{IMRPhenomXPHM} model: GW151226 \cite{GW151226, GW151226alternate} and GW190521 \cite{GW190521, GW190521alternate}. Both of these newly found solutions indicate a large spin misaligned with the orbital angular momentum on the primary mass and a low mass ratio $q < 1$. The GW151226 reanalysis showed that the new low-$q$ solution were absent from the posterior distribution of the source parameters when either the higher modes or the precession effects were neglected \cite{GW151226alternate}. The original analysis of GW190521 used the \texttt{IMRPhenomPv3HM} model \cite{IMRPhenomPv3HM}, which had higher modes and precession, but not the improved calibration of the subdominant modes, and particularly not in the low-$q$ regime \cite{GW190521, GW190521alternate}. The other two injections are artificially made to have a high total mass and large in-plane spins: one with a high mass ratio $q < 1$ and another with a low mass ratio. The parameters used for the injections are listed in \reftab{injectionparameters}.

\begin{table*}[!]
\caption{\label{tab:injectionparameters}
The parameters injected for the 4 injections used in this work, and the fiducial parameters for GW190814 found by using the global optimization algorithm \texttt{differential\_evolution} provided in \texttt{scipy}. We generated waveforms using the \texttt{IMRPhenomXPHM} model with these parameters and injected them at the 8 s mark of a 16 s stream centered at the GPS time collected at a sampling rate of 1024 Hz.}
\begin{ruledtabular}
\begin{tabular}{c|ccccc}
& High $q$ & Low $q$ & GW151226-like & GW190521-like & GW190814 (fiducial) \\
\hline
Primary mass $m_1$ [$M_\odot$] & 30.0 & 100.0 & 29.3 & 168.0 & 24.4 \\
Secondary mass $m_2$ [$M_\odot$] & 20.0 & 10.0 & 4.3 & 16.0 & 2.7 \\
$\chi_{1,x}$ & 0.700 & 0.700 & 0.600 & 0.641 & 0.024 \\
$\chi_{1,y}$ & 0.000 & 0.000 & -0.200 & 0.000 & 0.039 \\
$\chi_{1,z}$ & 0.000 & 0.000 & 0.500 & -0.559 & 0.002 \\
$\chi_{2,x}$ & 0.000 & 0.000 & 0.000 & 0.000 & -0.305 \\
$\chi_{2,y}$ & 0.700 & 0.700 & 0.000 & 0.000 & -0.574 \\
$\chi_{2,z}$ & 0.000 & 0.000 & 0.000 & 0.000 & -0.067 \\
Luminosity distance $D_L$ [Mpc] & 496.91 & 496.91 & 457.00 & 400.00 & 283.28 \\
Line of sight $\theta_{\text{JN}}$ [rad] & 1.200 & 1.200 & 0.403 & 0.500 & 2.408 \\
$\kappa$ \footnote{As defined in Appendix C of Ref.~\cite{twistagain}} & 0.200 & 0.200 & 0.000 & 0.000 & 2.730 \\
Right ascension $\alpha$ [rad] & 3.000 & 3.000 & 3.831 & 3.340 & 0.412 \\
Declination $\delta$ [rad] & 0.100 & 0.100 & -0.819 & 0.600 & -0.574 \\
Polarization angle $\psi$ [rad] & 0.500 & 0.500 & 2.698 & 0.000 & 2.962 \\
GPS time & 1242443867.4 & 1242443867.4 & 1135137260.6 & 1242443867.4 & 1249852257.0
\end{tabular}
\end{ruledtabular}
\end{table*}

To start with, we need posterior source parameters for which we can evaluate the log likelihood both approximately using our new mode-by-mode binning method and exactly. The samples are collected using \texttt{MultiNest} with $\texttt{nlive} = 2000$. For these runs, the log likelihoods are evaluated using the original relative binning method of Ref.~\cite{relativebinning}. We use the parameter choice $\epsilon=0.03$ for the maximum allowed differential phase (see Eq. 10 of Ref.~\cite{relativebinning} for the definition of $\epsilon$), which is sufficient for obtaining accurate posterior samples. For the injected events, the injected parameters are used to set the fiducial waveform. For the real GW190814 event, fiducial parameters are found using \texttt{scipy}'s \texttt{differential\_evolution} \cite{differentialevolution}, a global optimization routine to (approximately) find the maximum likelihood fit to the data. 

We obtain the strain data containing GW190814, as well as those used as the noisy background for our injections, from the public database GWOSC \cite{ligoopendata}. For each event, we collect a 16-second time series centered on the GPS time listed in \reftab{injectionparameters} with a sampling rate of 1024 Hz. The injections are added at the 8 second mark into the noisy 16-second time series. PSDs are measured using Welch's method from a 128-second time series centered at the same GPS time with overlapping 16 s segments. We only include frequencies from 20 Hz to 512 Hz in the likelihood evaluation, as there is negligible SNR contribution outside this frequency range. 

\begin{figure}[!]
\begin{tabular}{c}
  \includegraphics[width=85mm]{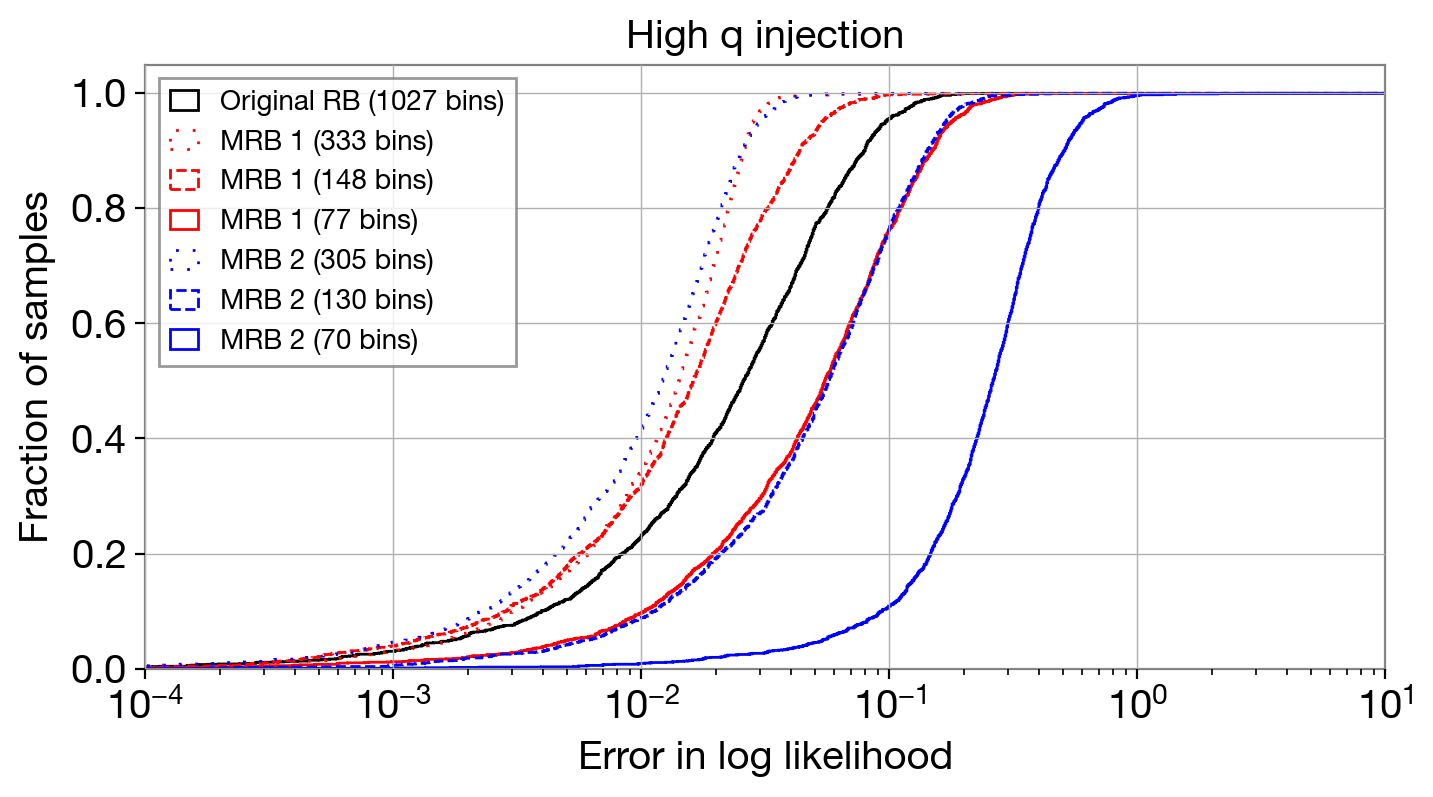} \\   \includegraphics[width=85mm]{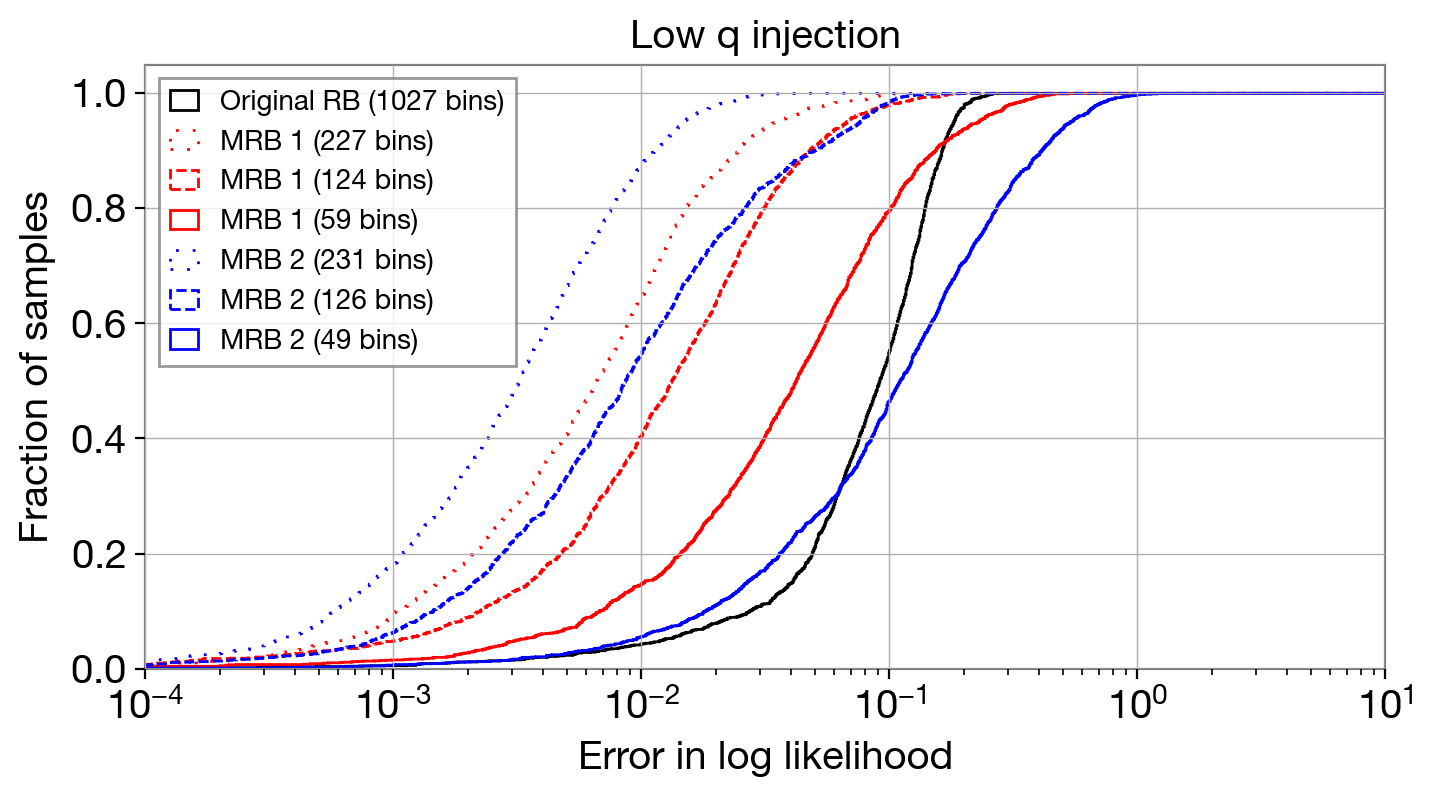}
\end{tabular}
\caption{CDFs for the {\it absolute} error in the log likelihood from the original relative binning and the mode-by-mode relative binning (MRB), evaluated for 2000 posterior samples collected for two synthetic signals we analyze. Different lines for the same scheme correspond to different values of $\eta$, which result in different numbers of bins. Both MRB schemes are able to outperform the original relative binning for the worst samples with fewer bins.}
\label{fig:moneyplots1}
\end{figure}

\begin{figure}[!]
\begin{tabular}{c}
 \includegraphics[width=85mm]{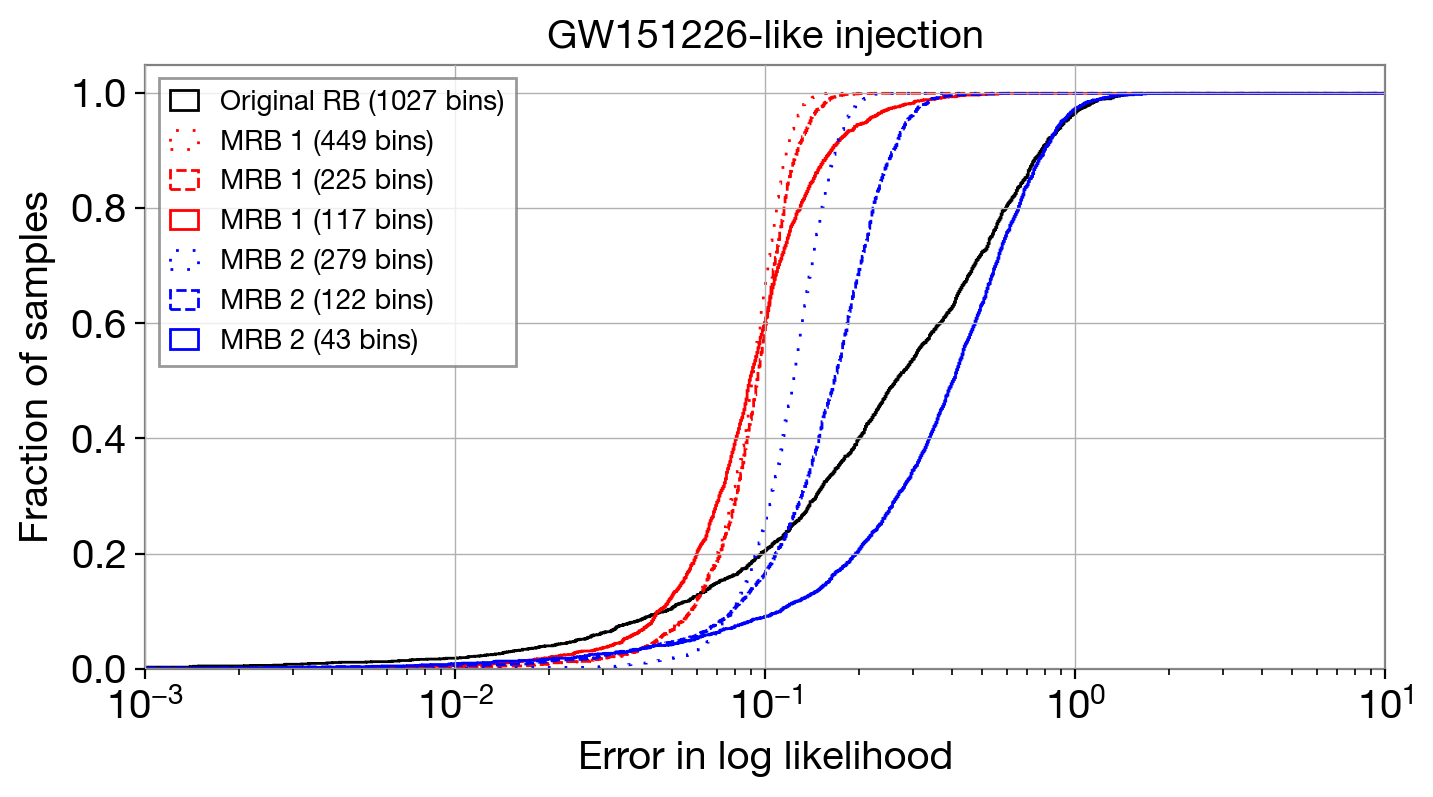} \\   \includegraphics[width=85mm]{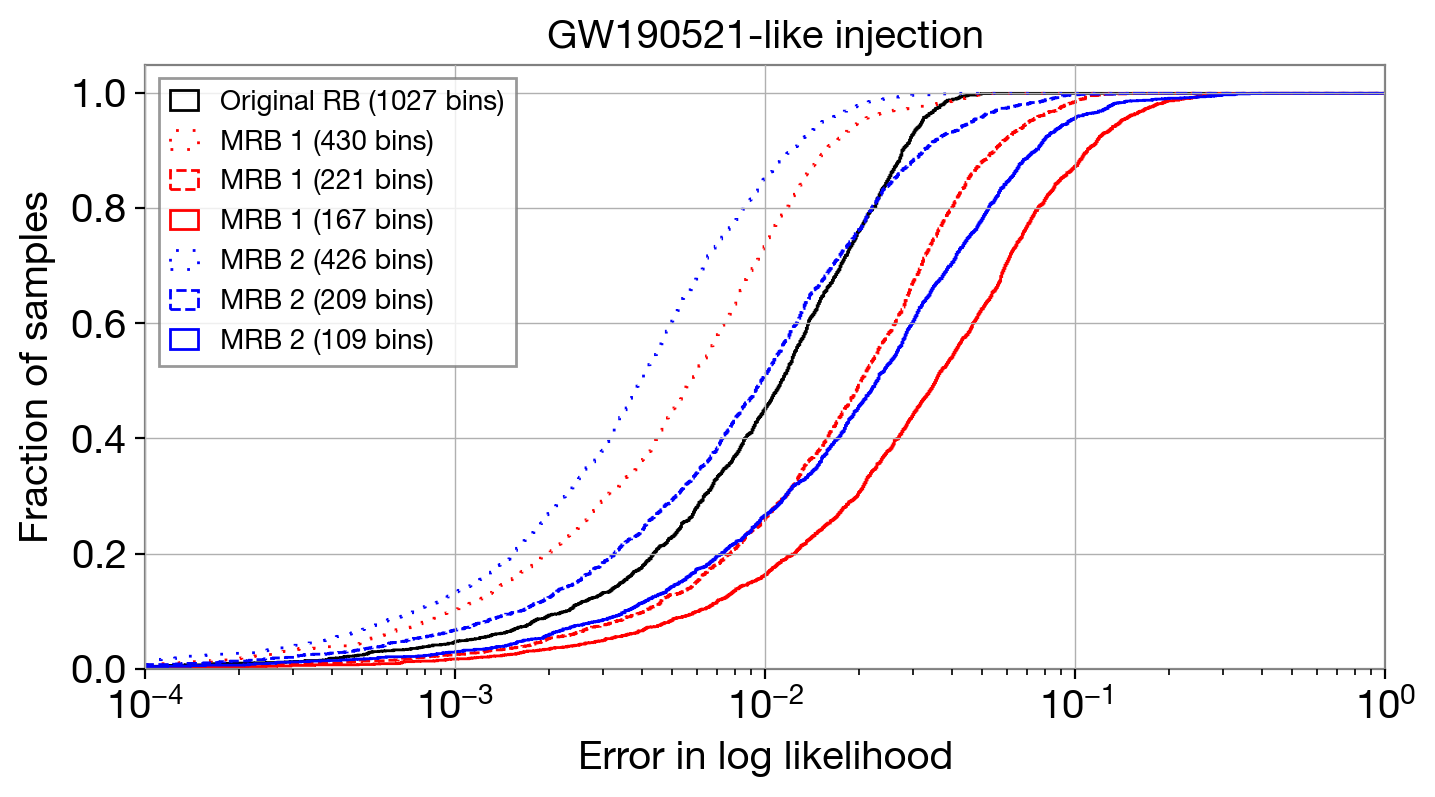} \\
 \includegraphics[width=85mm]{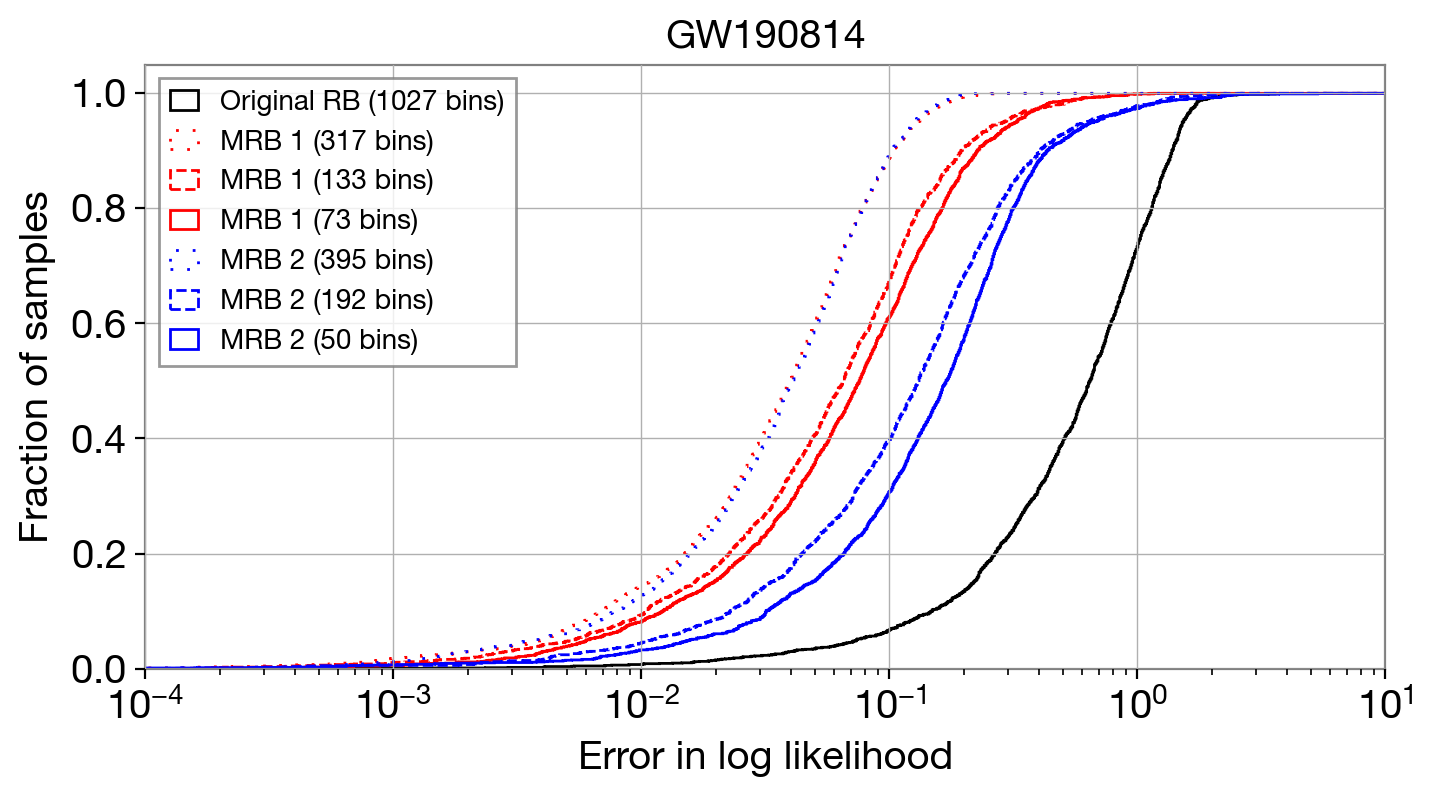}
\end{tabular}
\caption{
Same as \reffig{moneyplots1} but for the GW151226-like injection, the GW190521-like injection, and GW190814. We see better results from Scheme 1 in all test cases except the GW190521-like injection.}
\label{fig:moneyplots2}
\end{figure}

We make a comparison between the original relative binning method using the original bin selection algorithm (with the parameter $\epsilon=0.03$ as defined in Ref.~\cite{relativebinning}), and the mode-by-mode relative binning method using our new bin selection (for a range of $\eta$ values), evaluating for 2000 parameter combinations randomly drawn from the posterior distribution derived for each event or injection. The bin selection algorithm uses a random posterior sample selected as the test waveform. We study the absolute errors in the log likelihood computed for these samples, which are shown in \reffig{moneyplots1} and \reffig{moneyplots2}.

\reffig{moneyplots1} and \reffig{moneyplots2} show the cumulative distribution function (CDF) for the absolute error in the log likelihood for the 2000 posterior samples collected for each event. We compare the original relative binning method with mode-by-mode relative binning, testing both our approximation schemes and varying $\eta$. The varying numbers of bins result from using different $\eta$ values. The number of frequency bins is one less than the number of required frequency-domain waveform evaluations, which dominate the runtime, so the bin number is a good proxy for the runtime. 
In all examples we show, the original relative binning method uses the same number of frequency bins because the bin selection prescription only depends on the frequency range and the maximum differential phase $\epsilon$, which do not change from event to event.

In \refapp{targeterror}, we investigate the tolerable size of the absolute error in the log likelihood for obtaining accurate posterior distributions. Conservatively, we conclude that the log likelihood error should be no greater than $\sim 0.1$ as random errors with a standard deviation of $0.1$ appear to have unnoticeable effect in the posterior. 

For the high $q$ injection, and using only 148 bins with Scheme 1, the errors with mode-by-mode relative binning (MRB) are smaller than those with the original relative binning method using 1027 bins at all percentiles. Using about 300 bins, both Scheme 1 and Scheme 2 are able to achieve this. Notably, these three cases have the log likelihood error no greater than 0.1 for all samples. 

Similarly, for the low $q$ injection, the MRB error derived from about 125 bins strictly outperforms that of the original relative binning. The error distribution has a larger standard deviation than in the high $q$ case, but this is due to most of low error samples having even smaller errors. 

For all methods, the samples for the GW151226-like injection have larger errors, but using 117 bins with Scheme 1 the error is less than 0.1 for 60\% of the samples, and using more bins the spread of the worse errors is significantly reduced, and the error is nearly always less than 0.2 for roughly twice the number of bins. 

The GW190521-like injection is the only example studied here where Scheme 2 achieves lower errors than Scheme 1 does using comparable number of bins. Nevertheless, Scheme 1 achieves lower errors than the original relative binning scheme with 430 bins. For the GW190814 event, 90\% of the errors are less than 0.1, when using 300--400 bins with both Scheme 1 and Scheme 2. The errors are roughly a factor of two to four larger using only 73 bins under Scheme 1. In the 317 bin case, Scheme 1 reduces the error by nearly an order of magnitude at one third of the computational cost when compared with original relative binning. 

Overall, both MRB schemes achieve lower errors with fewer bins at all percentiles above 50\% when compared with the original relative binning method. In all cases, MRB achieves a median absolute error of 0.1 using substantially reduced number of bins. 

\section{Discussion}
\label{sec:discussion}

In all cases except the GW190521-like injection, Scheme 1 outperforms Scheme 2. We believe this is because the quantities linearized in Scheme 1 are usually smoother functions of the frequency than those linearized in Scheme 2, which reduces the required number of bins for a given accuracy goal. This can be seen in \reffig{smooth}. In fact, the ratio linearized in Scheme 2 is related to the one linearized in Scheme 1 through

\begin{eqnarray}
\left[r_{\ell, m'}(f)\right]_{\text{Scheme 2}} &= \frac{h_{\ell,m'}(f)}{h^0_{\ell,m'}(f)} = \frac{C_{\ell,m'}(f)\hat{h}_{\ell,m'}(f)}{C^0_{\ell,m'}(f)\hat{h}^0_{\ell,m'}(f)} \nonumber \\ 
&= \frac{C_{\ell,m'}(f)}{C^0_{\ell,m'}(f)}\left[r_{\ell, m'}(f)\right]_{\text{Scheme 1}}.
\label{eq:ratio2decomp}
\end{eqnarray}

The ratio linearized in Scheme 1 is similar to the ratio linearized in the original relative binning method applied to non-precessing waveform models. The $L$-frame mode components are equivalent to the $J$-frame components for a non-precessing waveform, and it has already been demonstrated that the ratio of non-precessing waveforms are smooth in the frequency space. However, the ratio $C_{\ell,m'}(f)/C^0_{\ell,m'}(f)$ can exhibit large oscillations or even sharp peaks when the denominator $C^0_{\ell,m'}(f)$ becomes small. This situation may arise because the coefficient $C_{\ell,m'}(f)$ is a linear combination of terms that may nearly cancel (see construction of $C_{\ell,m'}(f)$ in \refsec{twistingprocedure}).

\begin{figure}[!]
\begin{tabular}{c}
\includegraphics[width=85mm]{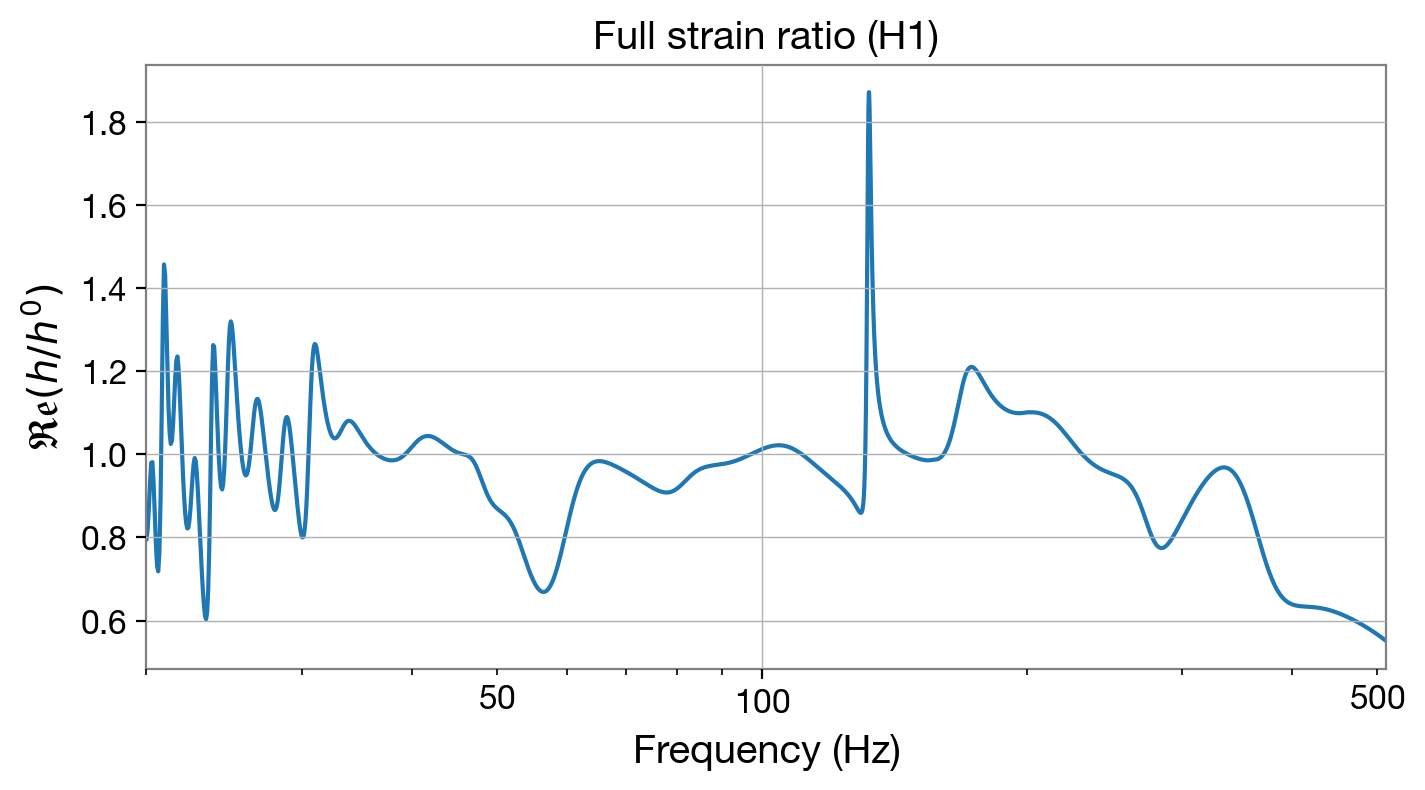} \\
\includegraphics[width=85mm]{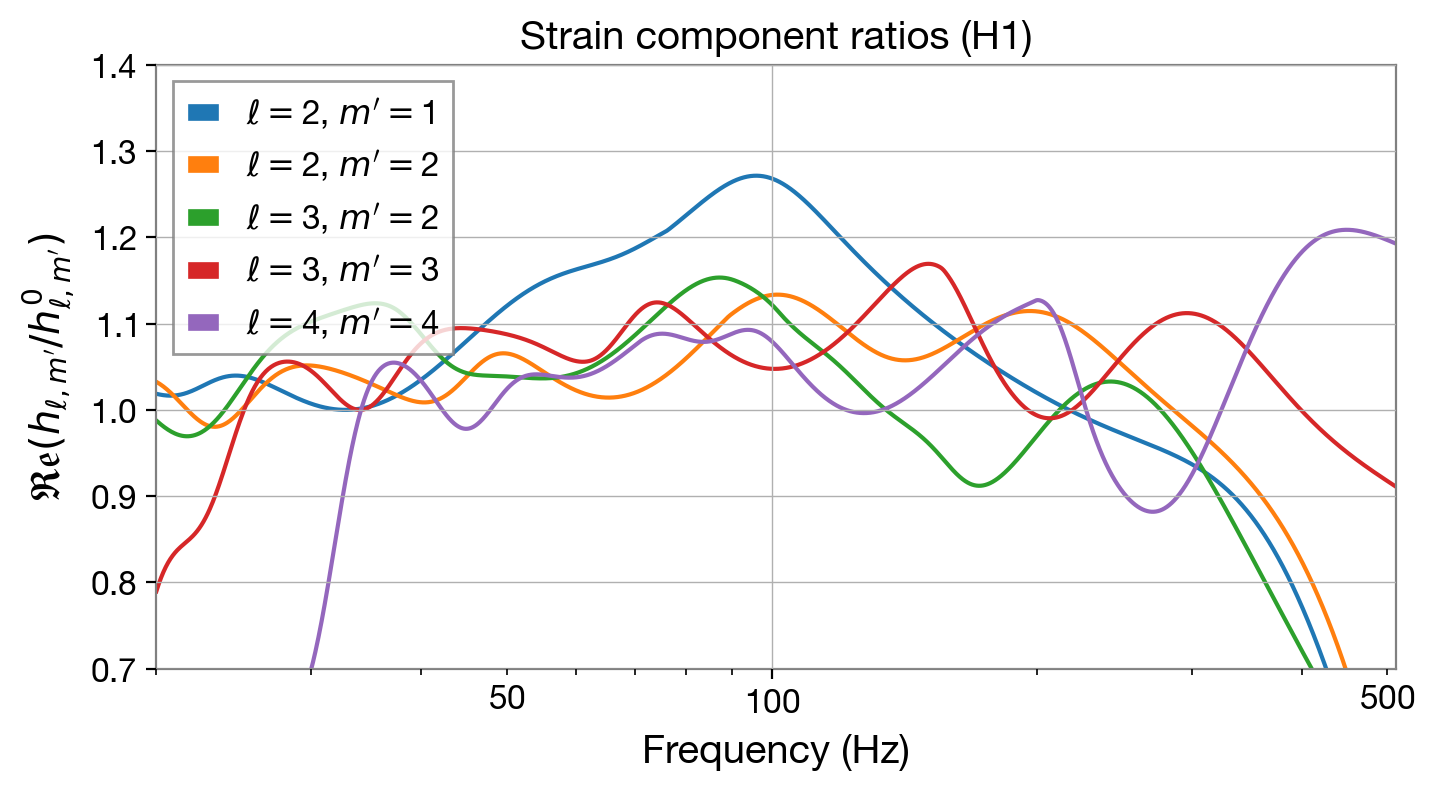} \\
\includegraphics[width=85mm]{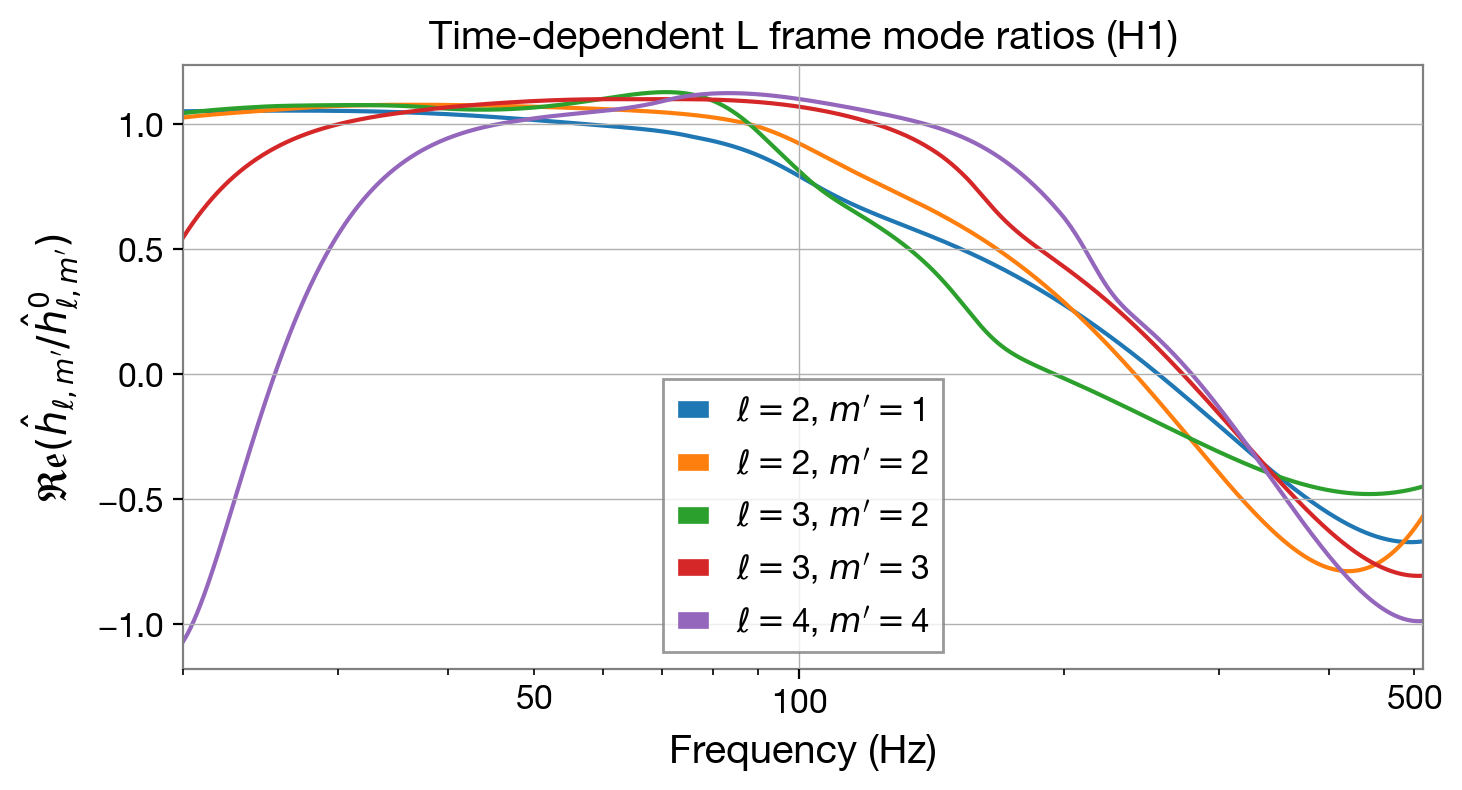} \\
\includegraphics[width=85mm]{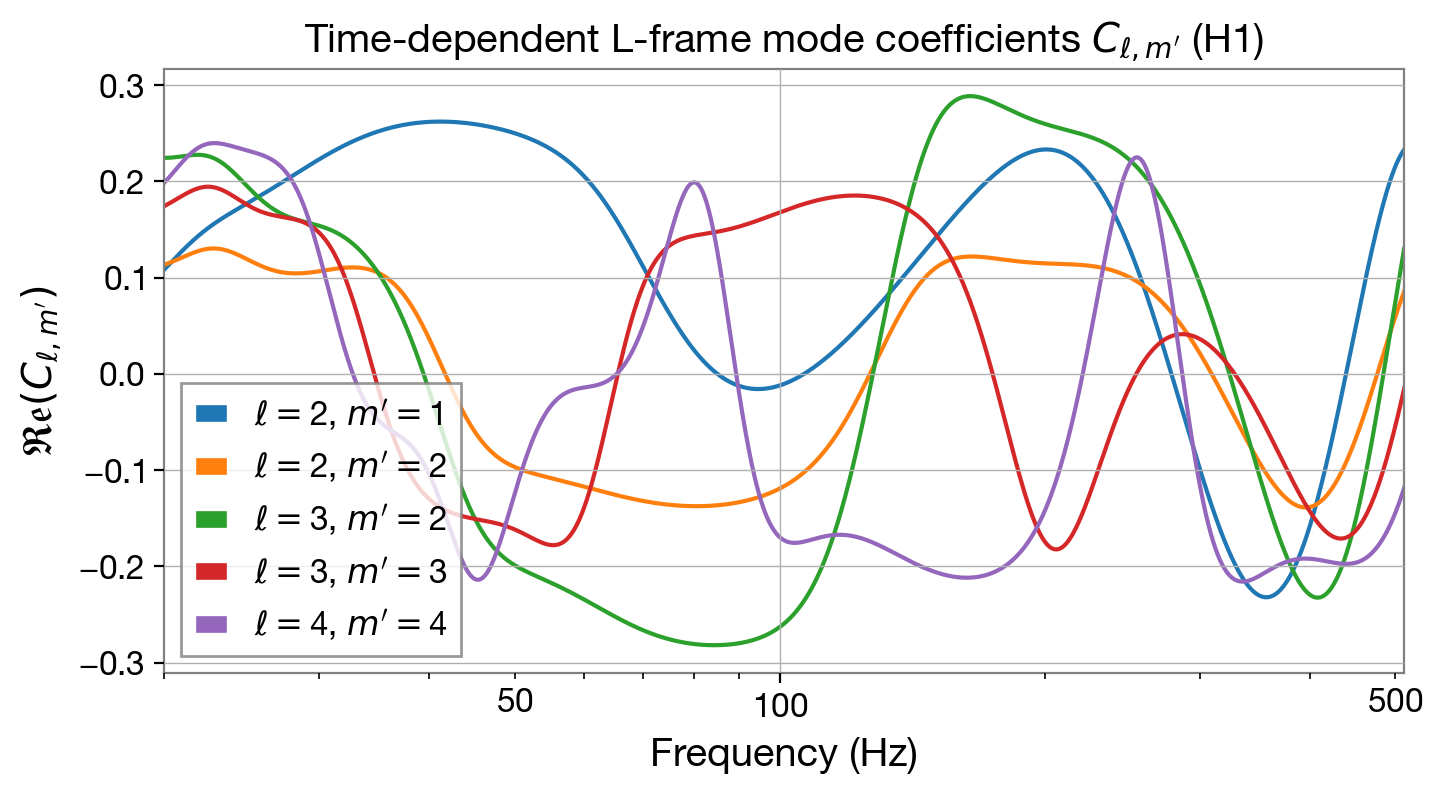}
\end{tabular}
\caption{
Same as \reffig{smooth} but for the GW190521-like injection. This is an example where Scheme 1 is not advantageous over Scheme 2. Again, mode-by-mode relative binning linearizes the time-dependent $L$-frame mode ratios and their coefficients (bottom two panels) under Scheme 1, and linearizes the strain component ratios (second panel) under Scheme 2.}
\label{fig:smooth2}
\end{figure}

However, it can also happen that Scheme 2 outperforms Scheme 1. When the aforementioned issue regarding $C^0_{\ell,m'}(f)$ does not arise, Scheme 2 may have an advantage because there are fewer quantities to be linearized so as to avoid compounding error. This appears to be the case in the GW190521-like injected event. The linearized quantities evaluated for one posterior sample of this event are compared in \reffig{smooth2}.
In this example, Scheme 1 is not advantageous over Scheme 2 in terms of frequency-space smoothness of the linearized quantities. 

Overall, we conclude that Scheme 1 is more robust because it accounts for the difficult case of small $C^0_{\ell,m'}(f)$, and more often yields better relative binning results, especially in the case of GW190814. 
In practice, the preferred scheme can be empirically determined alongside the process of choosing suitable fiducial and test waveforms.

To use mode-by-mode relative binning in practice, one has to set a fiducial waveform for computing the summary data, and a test waveform for determining the frequency bins. If the experimental collaboration has already provided full parameter inference results for the event under analysis, the reported parameters can be used as the fiducial parameters. If the signal under analysis is uncovered by a search pipeline, crude source masses and spins are typically reported, and can be used as the initial choice for the fiducial parameters. If not all parameters are reported, or are only reported for simplified waveform models, then the fiducial parameters can be obtained via global maximization of the exact likelihood. For well measured parameters, global optimization should be performed within the small vicinity of the reported values. For poorly measured parameters, optimization can performed over the entire physically allowed region.

If global optimization using the exact likelihood is computationally too slow, it is feasible to couple the relative binning approximation to the optimization. If the likelihood errors from using the initial choice of ficudial parameters are still too large, the fiducial parameters can be iteratively updated, just like what has been practiced using the original relative binning method~\cite{relativebinning, relativebinning170817}. As a matter of fact, there is great flexibility in setting the fiducial parameters, as many extrinsic parameters enter the individual modes only through frequency {\it independent} factors which do not degrade the accuracy of mode-by-mode relative binning.

Our results are minimally affected when changing which sample is used as the test waveform for the selection of frequency bins. This suggests that any set of test parameters in the support of the posterior distribution would be satisfactory. In practice, we recommend obtaining the test parameters by running a few iterations of a sampler starting from the fiducial parameters. The test parameters can be obtained by a few Metropolis-Hastings iterations, or a few steps of a more sophisticated sampler running on the exact log likelihood. 

\section{Conclusion}
\label{sec:concl}

In this paper, we address faster likelihood evaluations, one important aspect of speeding up gravitational wave parameter inference among several factors including sampling efficiency and parallel capabilities. We have developed the mode-by-mode relative binning method as an efficient approximation of the likelihood function, and have demonstrated its enhanced accuracy and theoretical speedup compared with the original relative binning method that does not take advantage of the multi-modal structure of the waveform.

The existing implementation can be found in our git repository at \url{https://github.com/nathaniel-leslie/modebymode-relative-binning}. With an optimized implementation applied to spin-orbit precessing waveforms including multiple harmonic modes, we expect the mode-by-mode relative binning method to run faster than comparable reduced order quadratures. An ROQ has been built for \texttt{IMRPhenomXPHM} using a code called PyROQ \cite{pyroq}. To evaluate the likelihood with ROQ, one must evaluate the linear and quadratic overlaps, which correspond to \refeq{Zdh} and \refeq{Zhh} respectively. Each overlap requires an interpolant to be evaluated, which requires the number of waveform evaluations that is equal to the number of basis functions for each interpolant. In Table III of \cite{pyroq}, it is shown that the total number of basis functions in their ROQ for \texttt{IMRPhenomXPHM} with a signal up to 1024 Hz ranged from 600 to 1100, depending on the parameter ranges. As shown in \reffig{moneyplots1} and \reffig{moneyplots2}, mode-by-mode relative binning can achieve the conservative target of absolute errors in the log likelihood error being around 0.1 on all test events with only around 100--200 frequency bins.

Another benefit of mode-by-mode relative binning is the ability to vary $\eta$ to flexibly trade speed for accuracy. This only requires recomputing bins by rerunning \refalg{binselection} with a new value of $\eta$, which takes a few seconds. If one achieves sufficient accuracy for a specific problem and can afford to sacrifice some, $\eta$ can be increased, which reduces bin count and runtime. Similarly, if one requires higher accuracy and can afford some additional runtime, $\eta$ can be decreased, which increases bin count and runtime but also improves accuracy.

Our conclusions about runtime are based on the reduction in the number of waveform evaluations, and nearly equivalently, the number of frequency bins. As we have mentioned previously, this is justified by the observation that waveform evaluation calls dominate the computational time of the likelihood evaluation. In the code used in this work, and for the \texttt{IMRPhenomXPHM} waveform model as an example, we have computed the $L$-frame modes $h^L_{\ell, m'}(f)$ and the Euler angles $\alpha(f)$, $\beta(f)$, and $\gamma(f)$ by calling the available \texttt{LALSuite} routines \cite{lalsuite}. These routines take approximately 3/4 of the time of a full likelihood evaluation using the full \texttt{LALSuite} routines for obtaining $h_+$ and $h_\times$. The remaining time is primarily spent on computing the twisting factors $C^+_{\ell, m'}$ and $C^\times_{\ell, m'}$, which involves a significant amount of evaluations of trigonometric functions and scales linearly with the bin number. Our current Python implementation of mode-by-mode relative binning is still inefficient due to overhead; it takes a few milliseconds to compute these twisting factors even for a few frequencies. The existing \texttt{LALSuite} routines for obtaining $h_+(f)$ and $h_\times(f)$ are sufficient for implementing the original relative binning method, which do not require computations of the twisting factors in Python. For this method, with 1027 bins, each likelihood evaluation takes roughly $\sim 4$-$5$ ms on a 2018 MacBook Pro with a 2.6 GHz Intel i7 processor. The time that it takes to evaluate the $L$-frame modes and Euler angles at the original bins is also about 3/4 of this time. We expect that an optimized routine coded in C, Julia, optimized Python packages, or in some other programming languages could realize such a performance. For the 133 bin evaluation shown in \reffig{moneyplots2}, the $L$-frame modes and Euler angles took approximately 0.7 ms. Assuming optimized implementation of the twisting factors, the same extrapolation suggests that sub-millisecond likelihood evaluations are achievable.

The fast evaluation times of the \texttt{LALSuite} code relative to our Python implementation demonstrate that the theoretical speed improvements of our code are realizable. In order to realize the full potential of mode by mode relative binning, this optimization must be performed. Once this is achieved, this method shows promise for substantially reducing the computational cost of parameter estimation for precessing waveforms with higher modes.

This method has only been tested on \texttt{IMRPhenomXPHM}, but we expect it to be similarly effective for other frequency-domain models with higher modes and precession. A simple application of this method to native time-domain models like \texttt{TEOBResumS} \cite{TEOBResumS} runs into a number of difficulties. Transforming these models into the frequency domain can be prohibitively slow, so one would hope to evaluate the likelihood in the time domain. In the time domain, the likelihood function becomes more than a simple frequency sum, as stationary noise produces a non-diagonal covariance matrix in the time domain. This particularly becomes difficult for relative binning with the $Z[h,h]$ term, as it would necessarily require sums over two time indices for each sampled waveform. New techniques will be needed to leverage the advantages of relative binning for these kinds of models. We are currently developing such relative binning methods for these time-domain models, and we hope to present tests of these methods in a following paper.

\acknowledgments

\section{Acknowledgements}
\label{sec:acknowledgements}
We would like to thank Uro\v{s} Seljak and Biwei Dai for useful discussions. NL is supported by N3AS under National Science Foundation grant PHY-1630782. LD acknowledges the research grant support from the Alfred P. Sloan Foundation (award number FG-2021-16495). GP is supported by STFC, the School of Physics and Astronomy at the University of Birmingham and the Institute for Gravitational Wave Astronomy.

This research has made use of data, software and/or web tools obtained from the Gravitational Wave Open Science Center (https://www.gw-openscience.org/ ), a service of LIGO Laboratory, the LIGO Scientific Collaboration and the Virgo Collaboration. LIGO Laboratory and Advanced LIGO are funded by the United States National Science Foundation (NSF) as well as the Science and Technology Facilities Council (STFC) of the United Kingdom, the Max-Planck-Society (MPS), and the State of Niedersachsen/Germany for support of the construction of Advanced LIGO and construction and operation of the GEO600 detector. Additional support for Advanced LIGO was provided by the Australian Research Council. Virgo is funded, through the European Gravitational Observatory (EGO), by the French Centre National de Recherche Scientifique (CNRS), the Italian Instituto Nazionale di Fisica Nucleare (INFN) and the Dutch Nikhef, with contributions by institutions from Belgium, Germany, Greece, Hungary, Ireland, Japan, Monaco, Poland, Portugal, Spain.

\appendix
\section{Target Log Likelihood Error}
\label{app:targeterror}

The absolute error in the log likelihood from mode-by-mode relative binning should be small enough not to bias the obtained posterior distribution for the source parameters. In this appendix, we study the impacts of log likelihood errors on the posterior distributions, using the high-$q$ injection as an example.

Our posterior samples have been obtained using the original relative binning method with finite likelihood errors. To obtain the exact (full FFT) posterior distribution, we use the method of importance sampling (IS). Let sample $i$ have the exact posterior 
\begin{equation}
    P_{\text{exact}}[i] = \frac{L_{\text{exact}}[i]\,p_i[i]}{Z},
\label{eq:exactposterior}
\end{equation}
where $L_{\text{exact}}[i]$ is the exact likelihood, $p[i]$ is the prior, and $Z$ is the evidence. Using the original relative binning method, we have
\begin{equation}
    P_{\text{ORB}}[i] = \frac{L_{\text{ORB}}[i]\,p_i[i]}{Z},
\label{eq:ORBposterior}
\end{equation}
where $L_{\text{ORB}}[i]$ is the relative binning likelihood. Importance sampling requires that samples are weighted by the ratio of the desired distribution over the sampled one, so to get samples of the exact posterior, each sample has the weight
\begin{equation}
    w[i] = \frac{P_{\text{exact}}[i]}{P_{\text{ORB}}[i]} = \frac{L_{\text{exact}}[i]}{L_{\text{ORB}}[i]} = e^{\ln L_{\text{exact}}[i] - \ln L_{\text{ORB}}[i]}.
\label{eq:ISweight1}
\end{equation}

We can also use importance sampling to simulate specific errors in the log likelihood. The distribution of the log likelihood errors due to either relative binning method is approximately Gaussian, so we can model these errors as Gaussian random variables. Let these errors be called $\delta[i]$; one is drawn for each sample from a Gaussian distribution. Then our simulated log likelihood with this error is 
\begin{equation}
    \ln L_{\text{sim}}[i] = \ln L_{\text{exact}}[i] + \delta[i].
\label{eq:simulatedlnL}
\end{equation}
So following \refeq{ISweight1}, the weight for the sample from this simulated distribution is 
\begin{equation}
    w_{\text{sim}}[i] = \frac{L_{\text{sim}}[i]}{L_{\text{ORB}}[i]} = e^{\ln L_{\text{exact}}[i] - \ln L_{\text{ORB}}[i] + \delta[i]}.
\label{eq:ISweight2}
\end{equation}

\begin{figure*}[!]
\includegraphics[width=75mm]{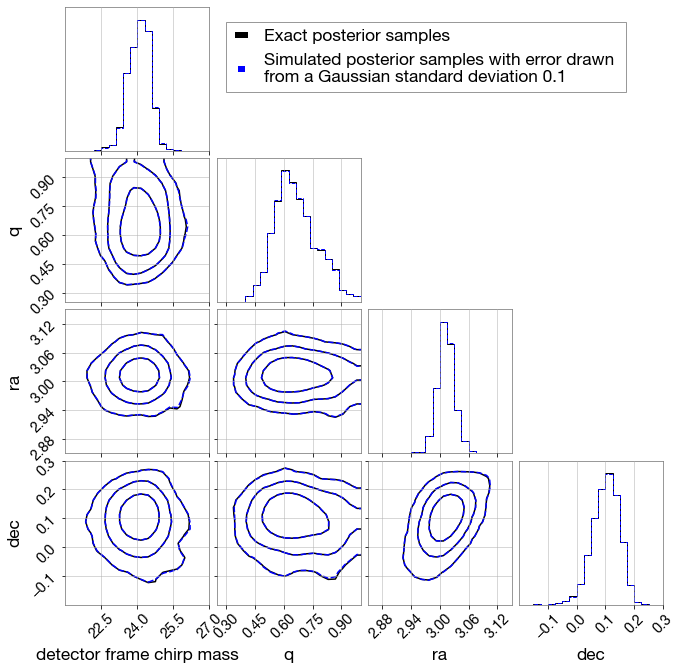} \includegraphics[width=75mm]{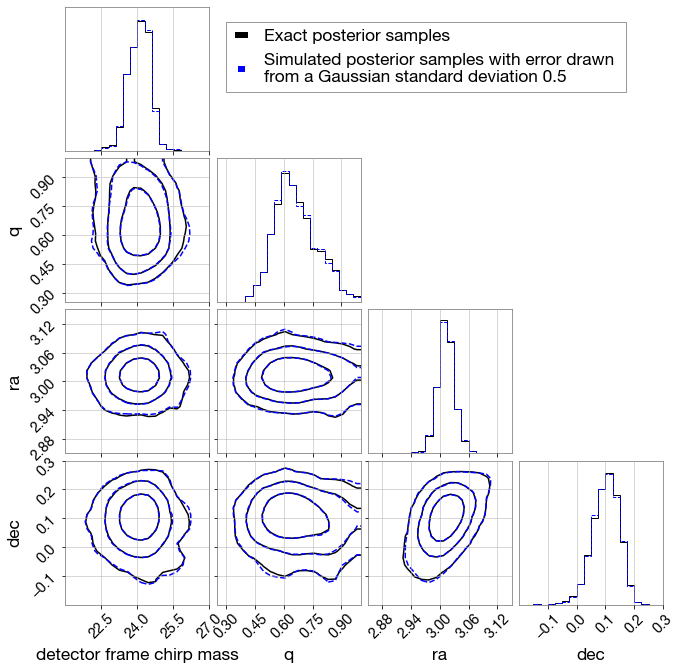} \\
\includegraphics[width=75mm]{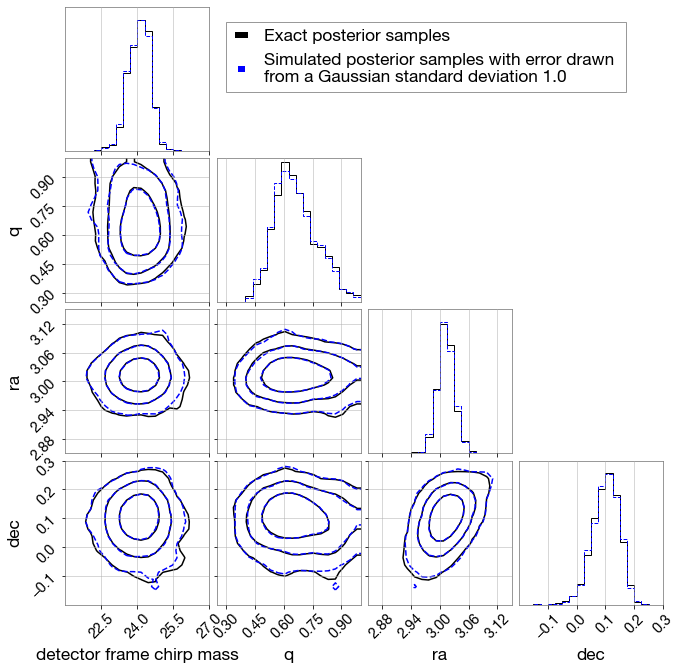} \includegraphics[width=75mm]{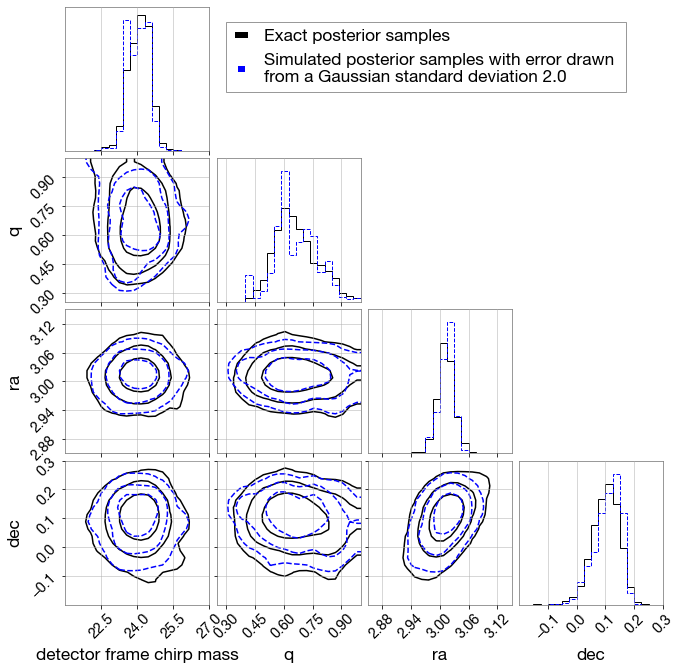} \\
\includegraphics[width=75mm]{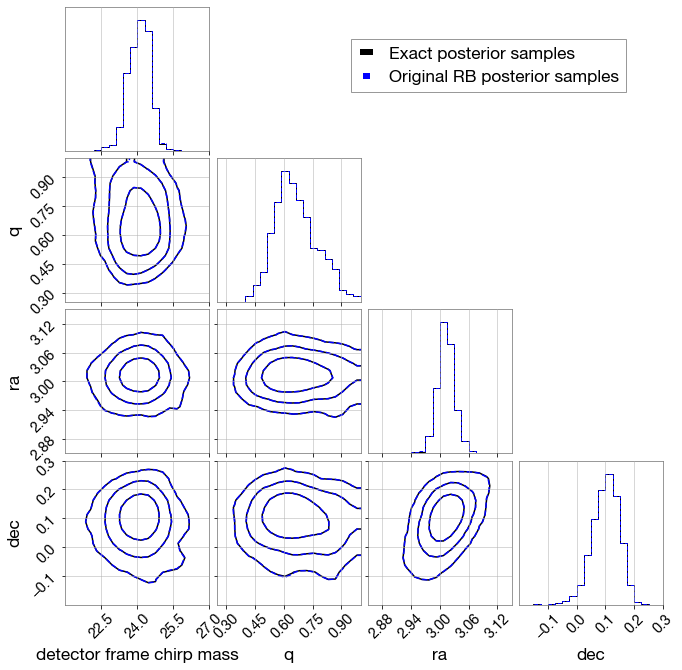} \includegraphics[width=75mm]{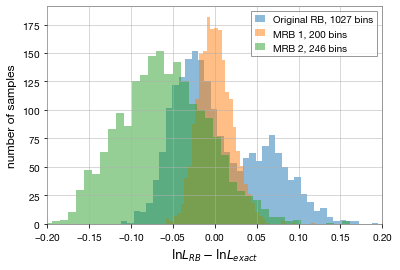}
\caption{The top 4 panels show the effect on the posterior distribution from zero-mean Gaussian random errors of various standard deviations in the log likelihood function. We apply the random errors to the high-$q$ injection as an example. The contours enclose 68\%, 95\%, and 99.7\% of the two-dimensional joint posterior samples. The bottom two panels concern actual errors from the various methods on the high q injection. The bottom left panel compares the posterior samples obtained from the original relative binning to the importance weighted posterior samples using the exact log likelihood. The bottom right panel shows the log likelihood errors for original relative binning, and our two relative binning schemes for the choice of $\eta=0.1$. The errors are not entirely dissimilar from Gaussian, and the original relative binning error shown makes no noticeable effect on the posterior distribution.}
\label{fig:iserror}
\end{figure*}

The above weights can be used to generate altered posteriors for comparison with an exact posterior. Such comparisons along with a histogram of the log likelihoods are shown for our high $q$ posterior in th top 4 panels of \reffig{iserror}. When the errors have a standard deviation of 0.1, they make virtually no visible impact in the posterior whatsoever. The simulated posteriors with log likelihood errors of standard deviation 0.5 and 1.0 have small inconsistencies, but their contours are generally unaltered in shape. However, once the standard deviation of the simulated error reaches 2.0, the contours are substantially distorted.

Additionally, in the bottom left of \reffig{iserror}, the actual posterior gathered from the original relative binning method and the importance weighted exact (full FFT) posterior do not differ at all. As can be seen in the blue histogram in the bottom right panel of \reffig{iserror}, the spread of the original relative binning errors log likelihood is a about 0.1.

In this example, we see that original relative binning error with a spread of about 0.1 is sufficiently small to not impact the posterior. Additionally the effect of the Gaussian error was negligible on the posterior at the same scale, and minimal up to an order of magnitude higher. Because this is only one event, we will be conservative, and use 0.1 as our target threshold, which is well justified by all of these results.

\bibliography{main}

\end{document}